\documentclass{aa}  

\usepackage{graphicx}
\usepackage{subcaption}
\usepackage{txfonts}

\newcommand{\vpeak}{v_{\rm peak}}
\newcommand{\vmax}{v_{\rm max}}
\newcommand{\mpeak}{m_{\rm peak}}
\newcommand{\msub}{m_{\rm sub}}
\newcommand{\minfall}{m_{\rm infall}}
\newcommand{\vinfall}{v_{\rm infall}}

\newcommand{\mvir}{m_{200\mathrm{c}}}
\newcommand{\rvir}{r_{200\mathrm{c}}}
\newcommand{\Mr}{{\rm M}_{r}}
\newcommand{\Mu}{{\rm M}_{U}}

\newcommand{\MTNG}{MTNG}
\newcommand{\MTNGdmo}{MTNG-DMO}

\newcommand{\Mstell}{m_{\rm stell}}
\newcommand{\Mgas}{m_{\rm gas}}

\newcommand{\hMsun}{ h^{-1}{\rm M_{ \odot}}}
\newcommand{\hMpc}{ h^{-1}{\rm Mpc}}

\newcommand{\sig}{\sigma_{8}}
\newcommand{\OmM}{\Omega_\mathrm{M}}
\newcommand{\Omb}{\Omega_{\rm b}}

\newcommand{\ns}{{n_{\rm s}}}

\begin{document}

   \title{The evolution of the baryonic content and mass profiles of satellite galaxies in the MTNG simulations}

   \author{Sergio Contreras
          \inst{1}\fnmsep\thanks{E-mail: scontreras1@us.es}
          \and
          Raul E. Angulo
          \inst{2,3}
           \and
           Giovanni Aricò
          \inst{4}
          \and
          Lurdes Ondaro-Mallea
          \inst{5,6}
          \and
        Sownak Bose
        \inst{7}
        \and
        Lars Hernquist
        \inst{8}
        \and        
        Ruediger Pakmor
        \inst{9}
        \and
        Volker Springel
        \inst{9}
          }

    \institute{
    Facultad de F\'isica, Universidad de Sevilla, Campus de Reina Mercedes, Av. Reina Mercedes s/n 41012 Seville, Spain
    \and
    Donostia International Physics Center, Manuel Lardizabal Ibilbidea, 4, 20018 Donostia, Gipuzkoa, Spain
         \and
    IKERBASQUE, Basque Foundation for Science, 48013, Bilbao, Spain
    \and
    INFN – Istituto Nazionale di Fisica Nucleare, Sezione di Bologna, 40127 Bologna, Italy
    \and
    Institute of Astronomy, University of Cambridge, Madingley Road, Cambridge CB3 0HA, United Kingdom
    \and
    Kavli Institute for Cosmology Cambridge, Madingley Road, Cambridge CB3 0HA, UK
    \and
    Institute for Computational Cosmology, Department of Physics, Durham University, South Road, Durham DH1 3LE, UK
    \and
        Harvard-Smithsonian Center for Astrophysics, 60 Garden St, Cambridge, MA 02138, USA
    \and
    Max Planck Institute for Astrophysics, Karl-Schwarzschild-Str. 1, D-85748 Garching, Germany
    }

   \date{Received March 14, 1592; accepted XXXX}
 
  \abstract
   {Empirical models often rely on key relations from the galaxy--halo connection to construct mock galaxy catalogues. These relations typically describe central galaxies (i.e. the main galaxy of a halo) more accurately than satellite galaxies, which are generally less massive and orbit within larger haloes. Satellite galaxies are affected by a variety of physical processes that pose significant challenges for modelling. In this work, we use \MTNG, a state-of-the-art cosmological  hydrodynamic simulation, to study the evolution of the baryonic component of satellite galaxies. Using the merger trees from this simulation, we follow the evolution of all $z=0$ satellite galaxies, tracking their stellar mass, gas mass, and $r$- and $U$-band magnitudes. We characterise this evolution using proxies including the fraction of subhalo mass and maximum circular velocity remaining relative to infall, the pericentric distance, and the time since infall. All of these quantities are commonly available in gravity-only simulations and can therefore be used to model these trends in simpler galaxy population models. We find that the gas mass, which is well described by the remaining subhalo mass fraction, declines much more rapidly than the other components, with satellites losing $\sim 80\%$ of their gas by the time the subhalo has lost half of its total mass. By contrast, the evolution of stellar mass and magnitudes is overall slower and is better described by the reduction of the host subhalo $v_{\rm max}$. We additionally provide an analytical model for the evolution of these properties. We then examine the evolution of satellite mass profiles. We find that, although stripping is strongest in the outer regions, the intermediate and inner parts of satellites experience mass loss at early times. The results of this work can be used by empirical models and galaxy formation models built on gravity-only simulations to improve their descriptions of satellite galaxies.}

   \maketitle
\section{Introduction}
\label{sec:intro}
The standard cosmological model, $\Lambda$CDM, predicts that dark matter haloes grow hierarchically, with larger haloes assembling largely through mergers with smaller haloes. Following a halo merger, the galaxies originally associated with the smaller haloes will eventually merge with the central galaxy of the larger halo (i.e. the galaxy at the centre of the halo, typically the most massive and brightest galaxy in the system) or be disrupted, although these events do not occur immediately. This scenario implies that massive haloes host multiple galaxies: one central galaxy and several satellite galaxies. During their infall into the host halo, and over successive orbits, satellite galaxies are affected by several physical processes, including starburst episodes, tidal stripping, strangulation, and gas stripping, which significantly alter their properties.

The properties of satellite galaxies, as well as their abundance and distribution within their host haloes, are the result of the cosmological model and the physics of galaxy formation. Galaxies form and evolve within dark matter haloes \citep{White:1978} until their host haloes merge with more massive systems, and the frequency of such halo mergers is set by the underlying cosmological model. Using N-body simulations, we have learned that the positions and velocities of satellite galaxies are well approximated by those of dark matter subhaloes in gravity-only simulations (i.e. dark matter-only simulations, whose evolution is determined solely by the cosmological model). This relation is one of the foundations of the galaxy--halo connection and has led to the development of the Subhalo Abundance Matching family of models, which are widely used to construct mock galaxy catalogues \citep{Vale:2006, Conroy:2006, ChavesMontero:2016, Contreras:2021_SHAMe}.

However, although the dynamics of galaxies are expected to be similar to those predicted by gravity-only simulations, they are not identical. In \cite{Contreras:2026a}, we studied the impact of baryons on the positions and velocities of dark matter subhaloes using the \MTNG\ hydrodynamic simulation \citep{MTNG_01,MTNG_02}, finding differences of a few per cent in the mean distance to the halo centre and in the velocity dispersion of satellite galaxies. Although this may appear to be a minor effect, \cite{Contreras:2026a} predict that it leads to 10--20\% variations in clustering amplitude, corresponding to 1--3$\sigma$ for DESI-like errors on small scales (i.e. below $1~\hMpc$). In addition, the detectability of satellite galaxies in observations depends on their baryonic content. As satellite galaxies lose their gas and undergo quenching, they can only be detected through observables linked to star formation if they were accreted recently, producing a significant detection bias towards the outskirts of their host dark matter haloes \citep{Orsi:2018}.

A good understanding of how the properties of satellite galaxies evolve is crucial for improving empirical models based on the galaxy--halo connection, such as the halo occupation distribution (HOD; \citealt{Jing:1998a, Benson:2000, Peacock:2000, Berlind:2003, Zheng:2005, Zheng:2007, Contreras:2013, Guo:2015a, Contreras:2017, Contreras:2023_HODev}) and subhalo abundance matching (SHAM). These models are widely used to characterise galaxy populations and to construct mock galaxy catalogues, which play a central role in the analysis and interpretation of galaxy surveys (e.g. \citealt{Rocher:2023, Yuan:2024, Ortega:2025}), survey design and preparation (e.g. \citealt{Smith:2022}), the construction of covariance matrices (e.g. \citealt{Manera:2013,Manera:2015,Lin:2020}), and even the inference of cosmological parameters (e.g. \citealt{Simha:2013,Contreras:2023_MTNG, Mahony:2026}; Ortega-Martinez et al., in prep.). The way satellite galaxies are modelled in these mocks can significantly impact the resulting clustering predictions, particularly in the non-linear regime, and therefore affect both cosmological and galaxy formation constraints derived from galaxy survey data.

In this work, we use the state-of-the-art hydrodynamic \MTNG\ simulation to study the evolution of the baryonic content of satellite galaxies. In the first part of this paper, we focus on characterising the evolution of their stellar mass, gas mass, and $r$- and $U$-band magnitudes. At both $z=0$ and $z=1$, we describe the evolution of these satellite properties in terms of subhalo properties, including subhalo mass loss, pericentric distance (defined as the minimum distance between the satellite and the host-halo centre), time since infall (defined as the time elapsed since a galaxy was classified as a satellite), and the remaining maximum circular velocity fraction. Some of these subhalo properties are directly available in the gravity-only simulations commonly used to construct mock galaxy catalogues (e.g. \citealt{Angulo:2021}), while others can be derived from their merger trees. We parametrise several of these relations to provide empirical models with a more accurate description of satellite galaxies. Some of these properties, particularly subhalo mass loss, have previously been linked to the evolution of satellite galaxies (e.g. \citealt{Penarrubia:2008, Smith:2013, Smith:2016, He:2026}) and have also been used in empirical models of satellite evolution (\citealt{Moster:2020}, using subhalo mass loss, and \citealt{Behroozi:2019}, using changes in $\vmax$). Nonetheless, a systematic study that explores the evolution of several galaxy properties as a function of multiple subhalo properties, while also assessing the performance and limitations of these approaches, remains absent from the literature.

In the second part of this paper, we investigate the evolution of the mass profiles of satellite galaxies during their infall into the host halo. We find that satellite galaxies lose their gas much more rapidly than their other mass components, including gas located in their central regions. We also find that, although stripping is strongest in the outer regions, the intermediate and inner parts of satellites experience mass loss at early times (i.e. before the galaxy is formally classified as a satellite).

The results of this work quantify and characterise the evolution of both the amount and the spatial distribution of mass in satellite galaxies, and complement those of \cite{Contreras:2026a}, which measure the impact of baryons on the positions and velocities of subhaloes. The trends identified in these studies can be incorporated into both empirical galaxy--halo connection models, such as HOD and SHAM-like models, and galaxy formation models built on gravity-only simulations, such as semi-analytical models (SAMs; \citealt{Kauffmann:1993, Cole:1994, Somerville:1999, Bower:2006, Lagos:2008, Somerville:2008, Benson:2010, Benson:2012, Jiang:2014, Croton:2016, Lagos:2018, Henriques:2020}) and other empirical approaches \citep{Yang:2003, Zheng:2009, Cacciato:2013, Guo:2016, Behroozi:2019, Moster:2020}, to improve their description of satellite galaxies. These results will also help to model the baryon distribution in the context of satellite galaxies, thereby affecting lensing statistics.

The outline of this paper is as follows. Section~\ref{sec:sims} presents the \MTNG\ simulation and describes the galaxy and subhalo samples used in this work. Section~\ref{sec:sat_ev} presents the evolution of the baryonic content of satellite galaxies, and Section~\ref{sec:mass_profile} shows the evolution of their mass profiles. We conclude in Section~\ref{sec:summary}. Appendix~\ref{sec:cen} presents a parameterisation of the gas content of central galaxies, which can be combined with the results of this paper to predict the gas content of satellite galaxies. Appendix~\ref{sec:cen_infall_sat} examines the intrinsic differences between central galaxies and satellites shortly after infall, while Appendix~\ref{sec:add_param} explores additional parameterisations of galaxy properties that may facilitate their modelling in gravity-only simulations. Finally, Appendix~\ref{sec:SUBFIND-HBT} provides an approximate assessment of the bias in satellite gas masses introduced by the version of SUBFIND-HBT used in MTNG through a comparison with TNG300.

Unless otherwise stated, masses are given in units of $\hMsun$ and distances in units of $\hMpc$. All logarithms are in base 10. The virial mass definition used in this work is $\mvir$, defined as the total mass of a halo enclosed within a sphere whose mean density is 200 times the critical density of the Universe. Throughout this work, the term satellite galaxy refers to the full subhalo, including the associated dark matter subhalo and full gas content, rather than only the central core composed mainly of stellar mass and cold gas.

\section{Numerical simulations and galaxy samples}
\label{sec:sims}

In this section, we present the cosmological hydrodynamic simulation used in this work, the MillenniumTNG (\MTNG) simulation (Section~\ref{sec:MTNG}), and describe the criteria used to select galaxies in each sample for comparison (Section~\ref{sec:target}).

\subsection{The \MTNG\ simulation}
\label{sec:MTNG}

To quantify the evolution of satellite galaxy properties, we require a hydrodynamic simulation with a large volume to ensure robust statistics, as well as sufficiently high resolution to resolve multiple satellites per halo even after they have been stripped of most of their mass. One of the few simulations currently available with these characteristics is the largest hydrodynamic run in the MillenniumTNG suite of simulations \citep{MTNG_01,MTNG_02,MTNG_03,MTNG_04,MTNG_05,MTNG_06,MTNG_07,MTNG_08,Contreras:2023_MTNG}. This suite comprises tens of gravity-only (usually referred to as dark-matter-only) and hydrodynamic simulations, together with light-cone galaxy catalogues generated using semi-analytical models. The project extends the well-known Millennium \citep{Springel:2005} and IllustrisTNG \citep{TNGa,TNGb,TNGc,TNGd,TNGe,Nelson:2019} projects in a direction that enables accurate studies of the galaxy--halo connection and of the impact of baryonic physics on clustering, particularly over much larger cosmological volumes than previously possible.

The main simulation used in this work is the largest hydrodynamic simulation in this suite, which we refer to as the \MTNG\ simulation. It has a periodic volume of $(500\ \hMpc)^3$ and contains $4320^3$ dark matter particles and an equal number of gas cells, implying an average gas-cell mass of $2.00\times10^7\,\hMsun$ and a dark matter particle mass of $1.12\times10^8\,\hMsun$. The simulation adopts a cosmology consistent with \cite{Planck:2015}\footnote{$\OmM = 0.3089$, $\Omb = 0.0486$, $\sig = 0.8159$, $\ns = 0.9667$, and $h = 0.6774$}, identical to that used in the IllustrisTNG project.

The initial conditions for \MTNG\ were generated by fixing the amplitudes of the initial power modes to their expected {\it rms} values \citep{Angulo:2016}, a procedure that substantially reduces the impact of cosmic variance, at least for second-order statistics. The simulation was evolved with the moving-mesh code \texttt{AREPO} \citep{AREPO}, which includes star formation, radiative cooling, the growth of supermassive black holes, and the associated feedback from both supernovae and active galactic nuclei. The haloes in this simulation were identified using a {\tt FRIENDS-OF-FRIENDS} (FoF) group finder and a modified version of the {\tt SUBFIND}-HBT substructure finder \citep{Gadget-4}, adapted from the {\tt Gadget4} code to {\tt AREPO}. {In this implementation of the substructure finder, gas-cell membership in satellite subhaloes is assigned using previous subhalo membership to find possible candidates, analogously to dark matter and star particles. Because gas cells in {\tt AREPO} are not strictly Lagrangian and can exchange mass across cell interfaces, this procedure can lead to a systematic underestimation of the diffuse gas mass associated with satellites. This limitation should be taken into account when interpreting the gas content of satellite subhaloes in MTNG.} 

Galaxies form as agglomerations of star particles whose properties can be measured directly from the simulation. The simulation includes 265 snapshots, spanning the range from $z=63$ (when the initial conditions were generated) to $z=0$. The large number of snapshots provides a unique opportunity to follow in detail the infall of satellite galaxies, during which the relevant physical processes occur on short timescales. However, only the group and subhalo catalogues are available for most snapshots, while particle data are accessible only for a limited number of key outputs. In this work, we use the particle data at $z=0$ to study the mass profiles of satellite galaxies.

In Appendix~\ref{sec:cen_infall_sat}, we match the satellite galaxies in the \MTNG\ simulation to the subhaloes of a gravity-only simulation from the MillenniumTNG suite with the same number of dark matter particles, cosmology, initial conditions, and volume, and run with the \texttt{Gadget-4} code \citep{Gadget-4}. We refer to this simulation as \MTNGdmo. To match the catalogues, we use the procedure described in \cite{Contreras:2026a}, which traces satellites back to the snapshot at which they were still centrals using the merger trees of the simulations, and then matches the objects based on their mass and position.

\subsection{Target galaxy samples}
\label{sec:target}

\begin{figure}
\includegraphics[width=0.42\textwidth]{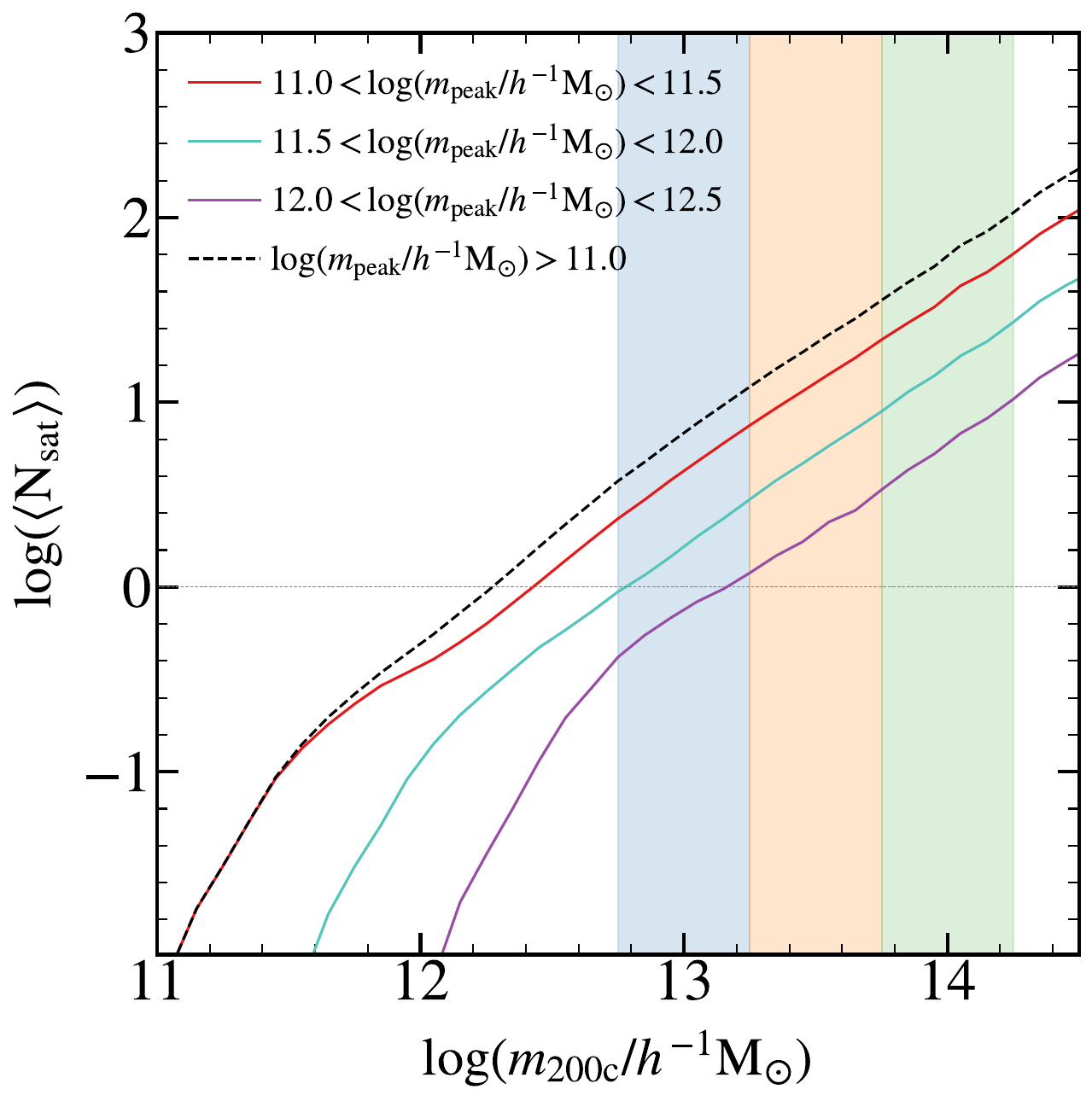}
\includegraphics[width=0.43\textwidth]{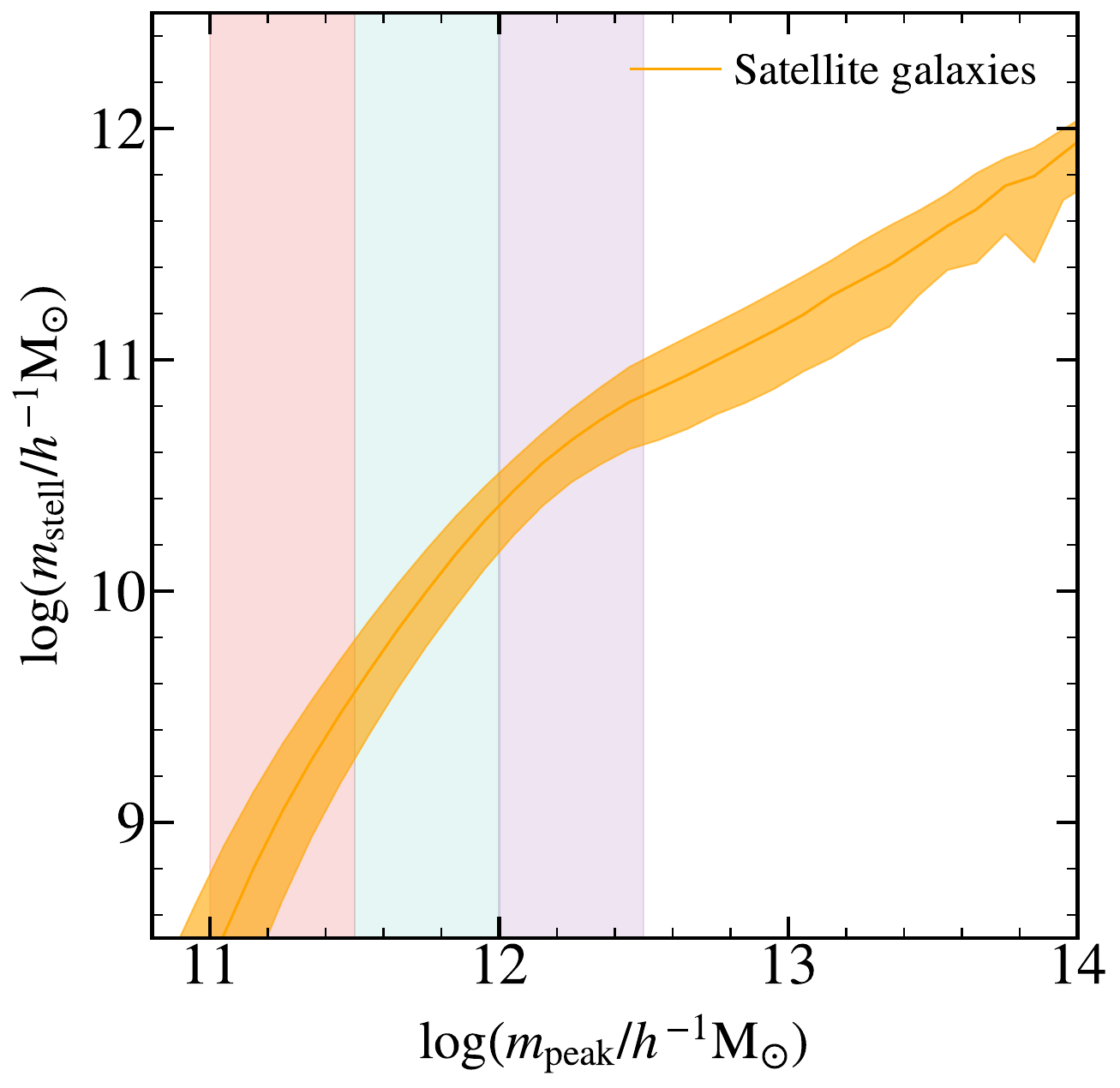}
\caption{(Top panel) Satellite halo occupation distribution for galaxies in the \MTNG\ simulation selected by their peak subhalo mass, i.e. the maximum subhalo mass reached by each galaxy during its evolutionary history. The solid lines represent galaxies selected by different peak subhalo masses, as labelled, while the dashed black line represents all subhaloes with peak subhalo mass above $10^{11}\ \hMsun$. The vertical coloured bands correspond to the halo-mass selection used throughout this work. (Bottom panel) Stellar mass of satellite galaxies as a function of peak subhalo mass. The solid line represents the median of the distribution, while the shaded regions represent the 16th and 84th percentiles. The vertical coloured bands correspond to the peak-subhalo-mass selection used throughout this work.}
\label{Fig:basic}
\end{figure}

In this work, we use satellite galaxies selected at $z=0$ and trace them back through the merger trees to $z\sim 13$. We study the evolution of four properties: stellar mass, gas mass, $r$-band magnitude, and $U$-band magnitude. Galaxies were selected within narrow ranges of host halo mass ($\mvir$), defined as the mass enclosed within a sphere whose mean density is 200 times the critical density of the Universe, and peak subhalo mass ($\mpeak$), defined as the maximum mass that a subhalo reaches during its evolutionary history. In particular, we select galaxies in three peak-subhalo-mass bins spanning $10^{11}-10^{12.5} \hMsun$, each 0.5 dex wide, and in three host-halo-mass bins spanning $10^{12.75}-10^{14.25} \hMsun$, also 0.5 dex wide. A visual representation of the selected samples is shown in the upper panel of Fig.~\ref{Fig:basic}. The coloured lines show the mean number of satellite galaxies per halo as a function of halo mass for the $\mpeak$-selected samples. The vertical shaded regions represent the additional halo-mass selection applied to the galaxy samples. Each sample should therefore be understood as the subset of the $\mpeak$-selected galaxies that also lies within the corresponding halo-mass bin.

The selection of galaxies as a function of $\mpeak$ is intended to avoid biasing the sample by the galaxies' evolutionary stage. As we will show in the next section, galaxy properties such as stellar mass or magnitudes evolve after infall. In addition, over most of the range considered, stellar mass is strongly correlated with $\mpeak$, as shown in the bottom panel of Fig.~\ref{Fig:basic}, which presents the stellar mass of satellite galaxies in the \MTNG\ simulation as a function of $\mpeak$.

\section{The evolution of satellite galaxies}
\label{sec:sat_ev}

\begin{figure*}
\includegraphics[width=0.95\textwidth]{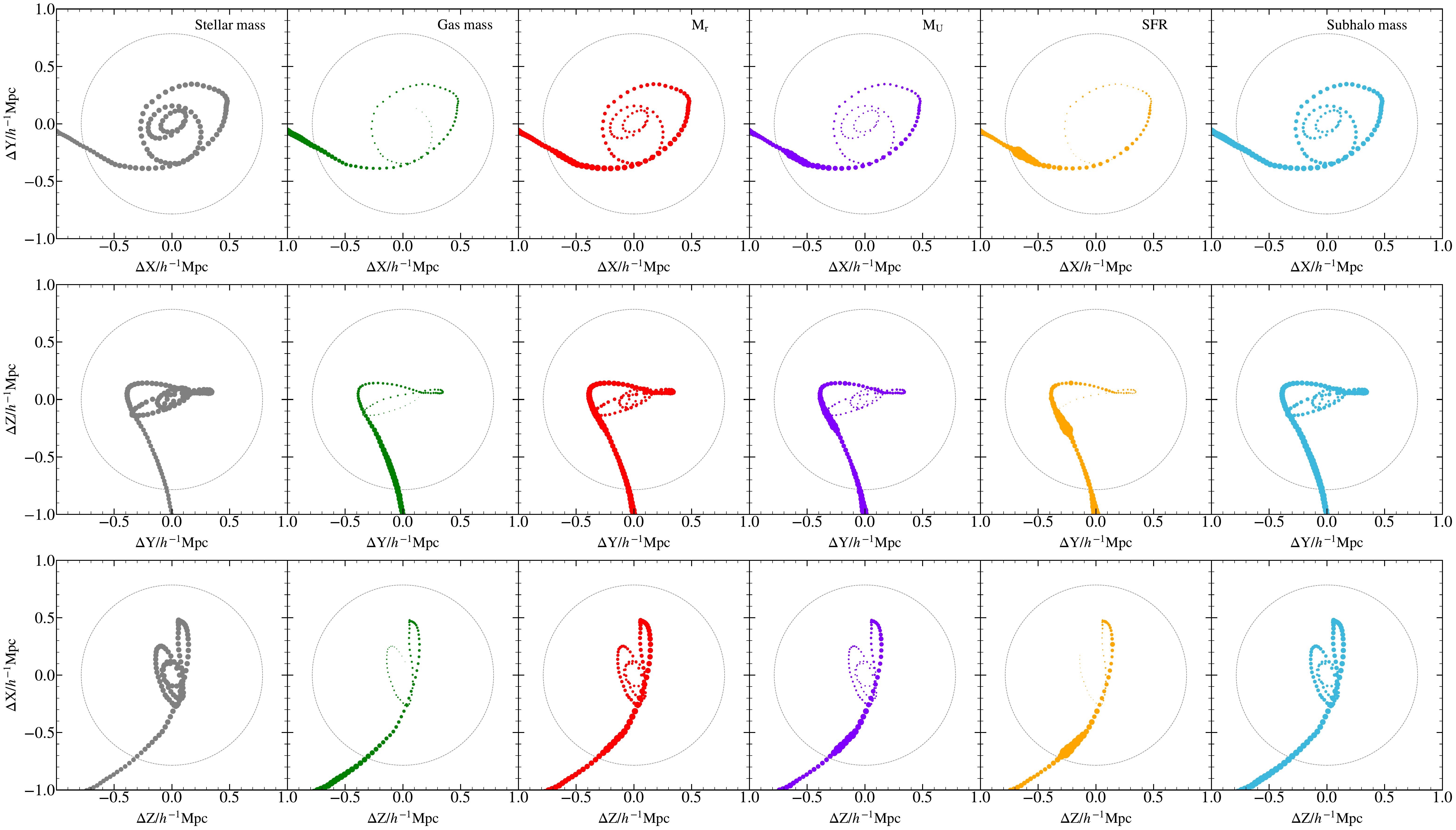}
\caption{Illustration of a satellite with $\Mstell \sim 10^{10.5}\ \hMsun$ infalling into a halo of $\sim 10^{13.5}\ \hMsun$ at $z=0$. The solid circles show the satellite's position relative to the halo centre. The size of each dot scales with the stellar mass, gas mass, star formation rate, and subhalo mass (in the first, second, fifth, and sixth columns, respectively), and with the difference in the $r$- and $U$-band magnitudes (in the third and fourth columns), relative to their values at the moment of infall. The dotted larger circle represents one virial radius ($\rvir$).}
\label{Fig:orbit}
\end{figure*}

\begin{figure*}
\includegraphics[width=0.95\textwidth]{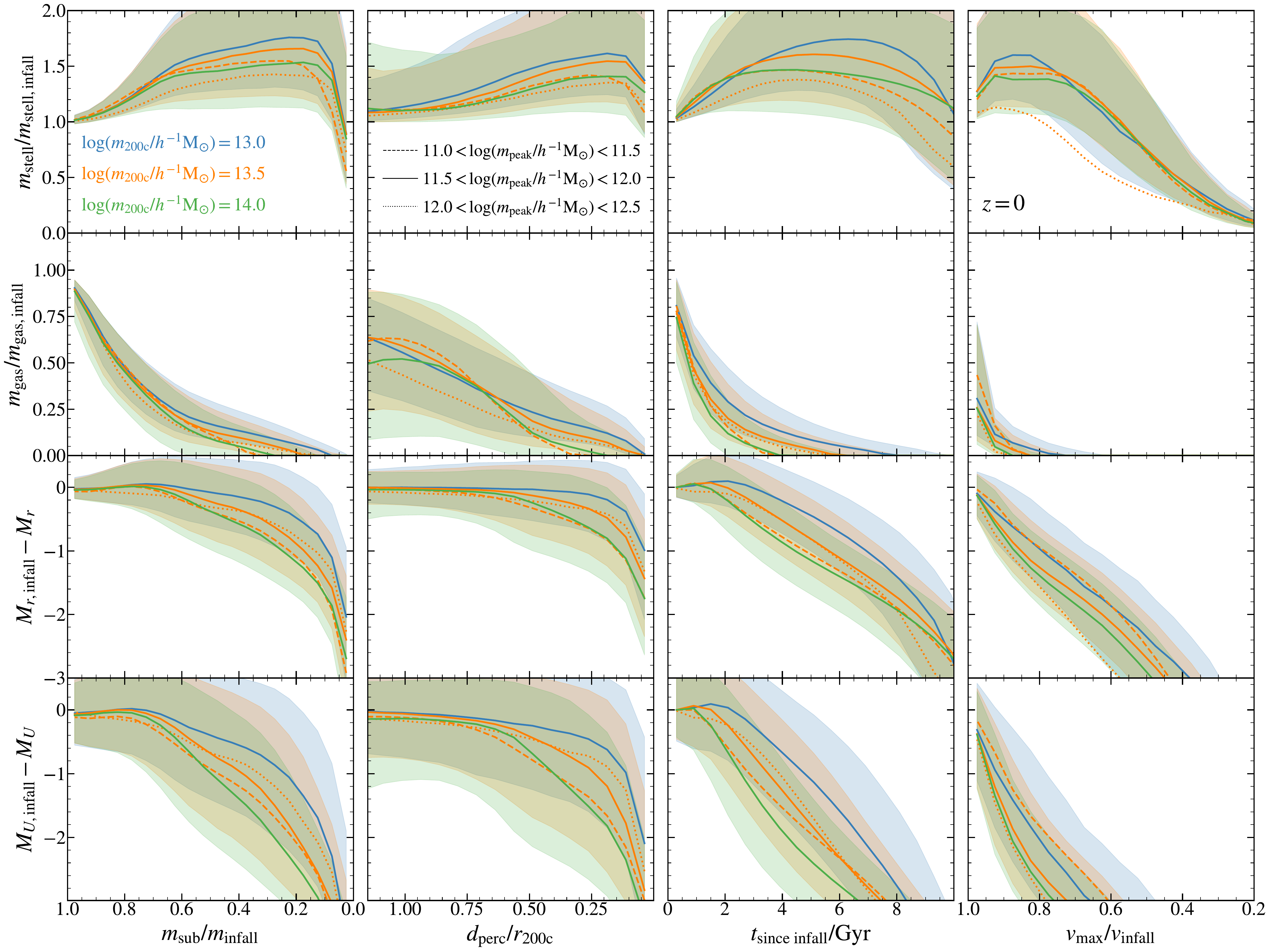}
\caption{The evolution of the stellar mass, gas mass, and $r$- and $U$-band magnitudes of satellite galaxies at $z=0$ is shown in the first, second, third, and fourth rows, respectively, as a function of the remaining subhalo mass fraction (first column), pericentric distance (second column), time since infall (third column), and the remaining maximum circular velocity fraction (fourth column). The different colours represent different host halo masses, while the different line styles represent different peak subhalo masses, as labelled. For simplicity, we show different halo masses only for the intermediate-$\mpeak$ sample. The shaded region represents the 16th and 84th percentiles of the distribution.}
\label{Fig:gen_ev}
\end{figure*}

\begin{figure*}
\includegraphics[width=0.95\textwidth]{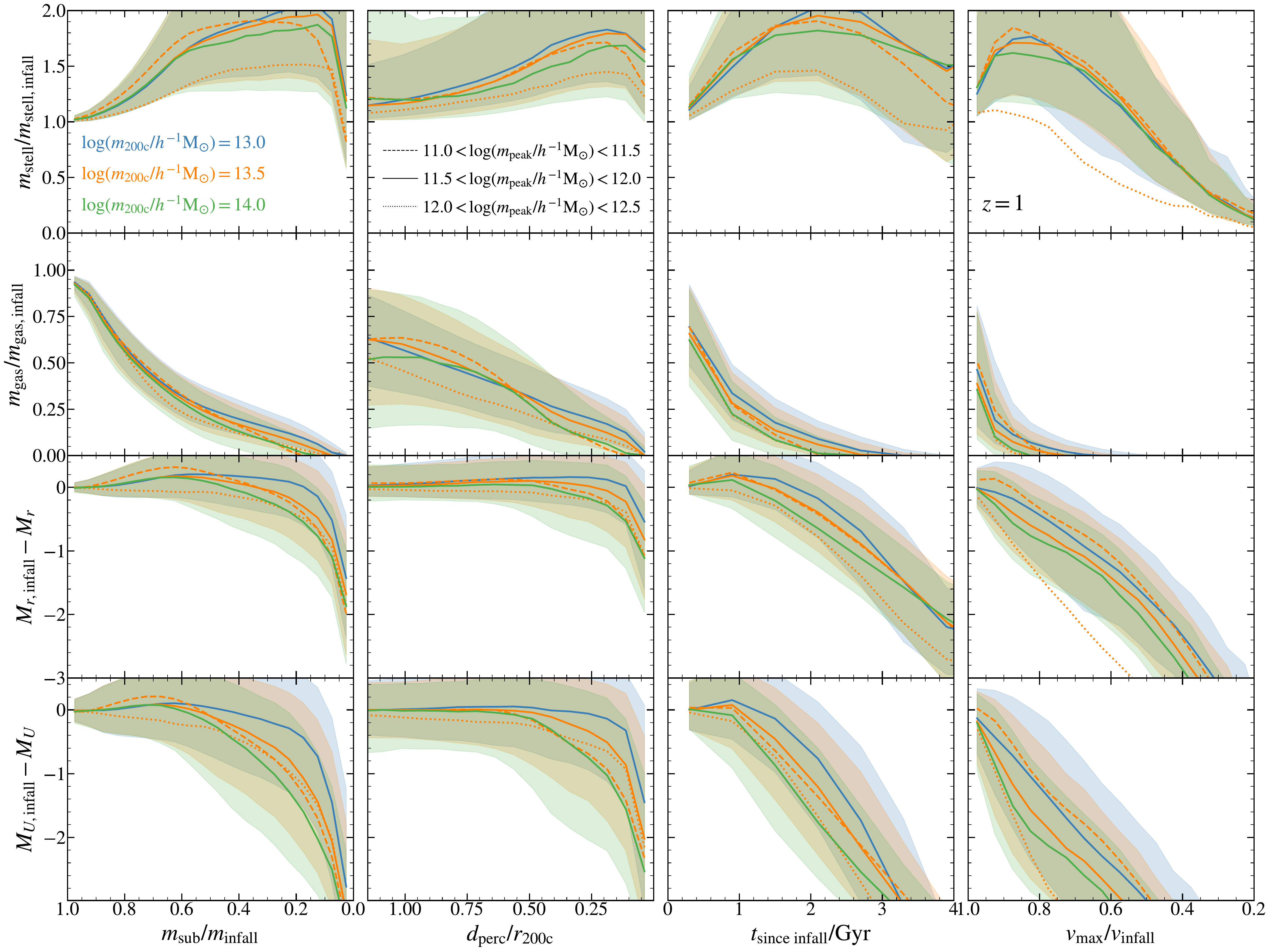}
\caption{Similar to Fig.~\ref{Fig:gen_ev}, but for $z=1$}
\label{Fig:gen_ev_z1}
\end{figure*}

\subsection{Evolution of the baryonic content in satellites}
\label{sec:EvBar}

We examine the evolution of the stellar mass, gas mass, and $r$- and $U$-band magnitudes of satellite galaxies. Figure~\ref{Fig:orbit} shows the evolution of these properties for a representative satellite galaxy with peak subhalo mass of $\mpeak \sim 10^{12.3}\ \hMsun$ orbiting within a halo of mass $\mvir \sim 10^{13.5}\ \hMsun$. In addition to the four previously mentioned properties, we also show the evolution of the star formation rate and the total subhalo mass of this galaxy, which are often used to characterise satellite evolution. Each dot represents the position of the satellite relative to the halo centre, which is located at the centre of each panel. Each column shows the evolution of a different property, as labelled, with the area of each dot representing the value of the corresponding quantity.
The figure shows that properties such as stellar mass and $\Mr$ change only modestly over most of the satellite's lifetime, with stronger evolution occurring at later stages. This is because stellar mass is concentrated in the inner regions of the satellite and is strongly gravitationally bound, which largely protects it from the stripping processes experienced by the subhalo as it moves through the host halo's hot gas and tidal field. It is expected that properties that are closely linked to stellar mass, such as red-band magnitudes, will also follow a similar evolutionary trend. By contrast, the gas and dark matter are less tightly bound and are more spatially extended, causing them to lose most of their mass during the first pericentric passages, when the tidal field experienced by the satellite is strongest (e.g. \citealt{Stucker:2023, Aguirre:2023}). The gas content declines more abruptly than the total mass, with virtually no gas left bound to the satellite before the galaxy completes its second orbit around the host. Finally, we note that the star formation rate also declines rapidly, but exhibits a strong starburst episode around the time of pericentric passage, clearly visible in the Y--Z plane (middle row). Here, ram-pressure compression could temporarily enhance star formation before quenching. This effect is also reflected in the evolution of $\Mu$, which shows a temporary decrease (i.e. an increase in luminosity), since the $U$ band is sensitive to both the stellar mass of the galaxy and its star formation rate.

We now aim to quantify the evolution of larger galaxy samples. To do so, we use the criteria described in Section~\ref{sec:target}, which select galaxies according to their host halo mass and peak subhalo mass. Figure~\ref{Fig:gen_ev} shows the evolution of the stellar mass, gas mass, and $r$- and $U$-band magnitudes (first, second, third, and fourth rows, respectively) as a function of four quantities: the ratio of the current subhalo mass to its infall value, the pericentric distance normalised by the comoving virial radius of the host halo, the time since infall, and the ratio of the current maximum circular velocity to its infall value. We define $\vmax \equiv \max\sqrt{GM(<r)/r}$, meaning that this quantity measures the total mass in the central regions of subhaloes and should therefore be interpreted as a subhalo property rather than a direct measure of galaxy kinematics. These quantities were chosen because they characterise different aspects of satellite evolution and are commonly available in both hydrodynamic and gravity-only simulations.

The remaining subhalo mass fraction can be readily obtained in simulations with merger trees and is expected to correlate with the gas mass of galaxies, since both the gas and the dark matter are spatially extended throughout the subhalo. This correspondence is not exact, however, because gas stripping is additionally affected by ram pressure. This quantity has been used to study the evolution of stellar mass in satellite galaxies (e.g. \citealt{Smith:2016, He:2026}) and to model it in semi-empirical models such as EMERGE \citep{Moster:2020} and in early versions of the empirical model SHAMe \citep{Contreras:2021_SHAMe}.

The pericentric distance, defined as the minimum distance of the galaxy from the halo centre, has been suggested to be related to episodes of rapid mass loss, both in total mass and gas mass, as well as to quenching processes, since tidal forces are expected to be strongest near pericentre (e.g. \citealt{Gnedin:1999, Wright:2022, Stucker:2023}). This picture is consistent with the evolution of the individual galaxy shown in Fig.~\ref{Fig:orbit}. To facilitate the stacking of many satellites, we normalise this distance by the comoving virial radius of the host halo. The pericentric distance is relatively straightforward to compute in any simulation with merger trees, but its value will always be an upper limit to the true value because of the finite number of snapshots, unless higher-order interpolation between snapshots is performed (for example, in spherical coordinates). This can become a significant limitation in simulations with poor temporal resolution, but not in \MTNG, where, as shown in Fig.~\ref{Fig:orbit}, the large number of snapshots makes this calculation sufficiently accurate.

The time since infall is perhaps the most direct way to characterise the evolution of a satellite galaxy. Although its precision also depends on the temporal resolution, the associated error is easier to estimate than that of the pericentric distance, which is sensitive to the orbital speed of the satellite galaxy. This quantity is also used by the latest version of the empirical model SHAMe to model the fading of satellite galaxies \citep{Contreras:2023_LIL}.

The final quantity we use to characterise satellite evolution is the remaining maximum circular velocity fraction ($\vmax / v_{{\rm max},\rm infall}$). It is well known that $\vmax$ is strongly correlated with stellar mass, and for this reason it is commonly used in the construction of mock catalogues within the SHAM framework (e.g. \citealt{ChavesMontero:2016}). Unlike the subhalo mass fraction, $\vmax$ is primarily sensitive to the central regions of the subhalo and therefore requires higher resolution to be measured accurately, since a larger number of particles per subhalo is needed to achieve good precision. This quantity is used by semi-empirical models such as UniverseMachine \citep{Behroozi:2019} to describe the evolution of satellite galaxies. An additional limitation of using $\vmax$ as a proxy for satellite evolution is that, in hydrodynamic simulations, stellar particles eventually contribute to the calculation of $\vmax$, artificially enhancing part of the correlation between these two properties. As has been shown when matching hydrodynamic and gravity-only simulations, a tight correlation between these quantities is expected \citep{ChavesMontero:2016}. Moreover, as shown in Fig.~\ref{Fig:gen_ev}, the stellar mass of galaxies does not follow a monotonic relation with $\vmax$, so we do not expect this contamination to dominate the trends presented here. Nevertheless, we warn the reader that this is an additional limitation of using $\vmax$.

The way galaxy properties evolve varies from case to case. The stellar mass of satellite galaxies increases significantly after infall, by around 30\%--50\%, depending on both the mass of the original galaxy and that of the host halo. By contrast, the gas mass decreases much more rapidly than the stellar mass changes. This behaviour is not captured when the evolution is parameterised in terms of $\vmax$, since a large fraction of the gas is lost while $\vmax$ remains nearly constant, thereby washing out the trend. The magnitudes in both bands evolve in all the cases we consider and follow broadly similar trends to each other, although the $U$-band magnitude fades more rapidly (i.e. $\Mu$ increases faster) than the $r$-band magnitude.

We find that all the subhalo properties considered can reasonably capture the evolution of the galaxy properties studied. That said, some appear to be more suitable than others for describing the evolution of specific galaxy properties. In the case of stellar mass, only the change in $\vmax$ is able to trace the decrease in stellar mass down to values below 50\% of those at infall, while doing so with only weak dependence on galaxy sample and host halo mass. As mentioned above, this quantity is strongly correlated with a galaxy's total stellar mass, as it is determined by the mass distribution in the subhalo's central regions. When the stellar mass evolution is characterised by this subhalo property, we find only a weak dependence on host halo mass, which facilitates the use of these trends in models of satellite galaxy evolution. A mild dependence on host halo mass is seen only for the lowest-$\mpeak$ sample, where the trend deviates for galaxies in the most massive host haloes. In this subhalo sample, galaxies can contain as few as 20 stellar particles ($\sim 4\times10^{8}\hMsun$), meaning that the results are strongly affected by resolution limits. As discussed above, the use of $\vmax$ can itself be limited by the resolution of the simulation, which may explain why this disagreement is less evident when the evolution is characterised in terms of the other subhalo properties. The other samples with the same $\mpeak$ but different host halo masses show similar results. For this reason, we do not investigate this difference further, but nevertheless show the results to make the reader aware of the limitations of using $\vmax$ as a proxy for galaxy evolution. In addition to the maximum circular velocity, changes in the subhalo mass also capture the evolution of stellar mass, although less smoothly than $\vmax$.

For the gas mass, we find that the best way to characterise its evolution is by using the remaining subhalo mass fraction. When expressed as a function of this property, the relation shows only a small scatter and a weak dependence on the galaxy sample chosen. These characteristics make the remaining subhalo mass fraction an ideal candidate for modelling this galaxy property. Small differences between galaxy samples appear only once the galaxy has lost more than 80\% of its gas mass. Another subhalo property capable of characterising this evolution is the time since infall, although it shows larger scatter and a stronger dependence on the galaxy sample. We do not find that either pericentric distance or $\vmax$ is able to capture the evolution of this galaxy property. In the case of pericentric distance, the evolution of individual galaxies shows that satellites begin to lose gas soon after becoming satellites. Since galaxies become satellites at different virial radii, owing to differences in the shape and mass of the host halo, it is difficult to model a single overall trend for all galaxies. This limitation, however, does not preclude the usefulness of pericentric distance for modelling the evolution of satellite properties, but rather indicates that it becomes more difficult to characterise a heterogeneous galaxy sample drawn from different host haloes. In the case of $\vmax$, changes in this quantity occur only after galaxies have already lost a considerable amount of gas and total mass, which makes it unreliable as a proxy for modelling the evolution of this property. We also examine the evolution of the total baryonic content (stellar mass plus gas mass), which evolves similarly to gas mass but with more universal trends (i.e. no significant differences between galaxy samples) when expressed as a function of subhalo mass (not shown here). {As discussed in Appendix~\ref{sec:SUBFIND-HBT}, the diffuse gas mass assigned to satellite subhaloes in MTNG may be systematically underestimated. The absolute gas-mass scale should therefore be interpreted with caution. Nevertheless, comparison with TNG300 \citep{TNGa,TNGb} suggests that the overall trend is preserved, so this relation still provides a useful description of the relative evolution of satellite gas content.}

All the subhalo properties considered can describe the evolution of the magnitudes. When expressed as a function of the remaining subhalo mass fraction, the magnitudes show the greatest scatter among the different parameterisations, but still provide a good description of their evolution, with galaxies becoming gradually fainter as the mass fraction decreases. When expressed as a function of pericentric distance, the magnitudes remain roughly constant out to about half the comoving virial radius and then begin to evolve smoothly. When expressed as a function of time since infall, galaxies tend to maintain their magnitudes for 1--2 Gyr and then become fainter, with their magnitudes changing linearly with time. Finally, when expressed in terms of the $\vmax$ ratio, galaxies exhibit almost linear evolution with low scatter and show no major differences across different galaxy samples. Although the evolution of $\Mr$ and $\Mu$ differs in detail, their evolutionary trends are similar, which facilitates their modelling. 

We examine the evolution in other photometric bands and find qualitatively similar trends. Bluer bands are more sensitive to recent star formation, as blue, luminous, high-mass stars evolve off the main sequence much more rapidly than their redder, fainter, lower-mass counterparts. As the gas content declines and star formation is suppressed, galaxies become progressively fainter and redder, giving rise to the trends identified in this work. A more detailed discussion is provided by \cite{Chaves:2020,Chaves:2021}.

We also examine the evolution of satellite properties at $z=1$ (Fig.~\ref{Fig:gen_ev_z1}). We find trends similar to those at $z=0$, with the main exception of those expressed as a function of time since infall. In addition, the stellar mass tends to increase slightly more at $z=1$ than at $z=0$. Despite this difference, the overall agreement between the two redshifts suggests that these trends may be largely universal and could therefore be used in models of satellite evolution independently of the infall redshift of the galaxies.

Regarding the scatter in these relations, as discussed in Appendix~\ref{sec:cen_infall_sat}, the stellar and gas masses of satellite galaxies are correlated with the large-scale environment, which likely contributes to the scatter in these relations. This scatter is also expected to depend on other secondary halo properties, such as concentration or halo age, which may influence the evolution of both central and satellite galaxies (e.g. \citealt{Zehavi:2018, Artale:2018}). A more detailed analysis of the physical origin of this scatter could further improve our understanding of satellite evolution and will be addressed in future work.

\subsection{Modelling the evolution of satellite galaxies}
\label{sec:EvBar_fit}

\begin{figure*}
    \centering

    \begin{subfigure}{0.40\textwidth}
        \centering
        \includegraphics[width=\linewidth]{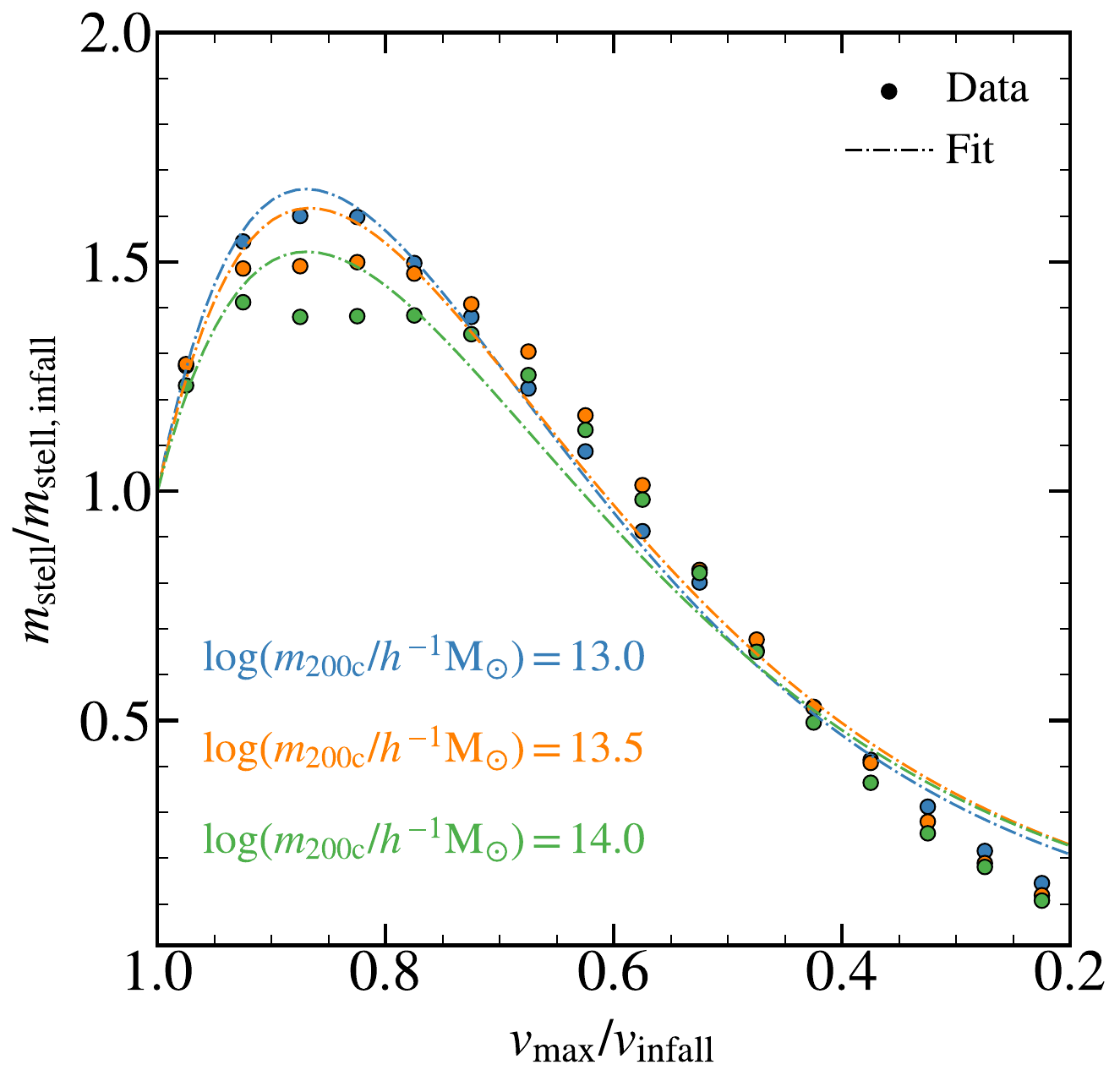}
    \end{subfigure}
    \hspace{0.03\textwidth}
    \begin{subfigure}{0.40\textwidth}
        \centering
        \includegraphics[width=\linewidth]{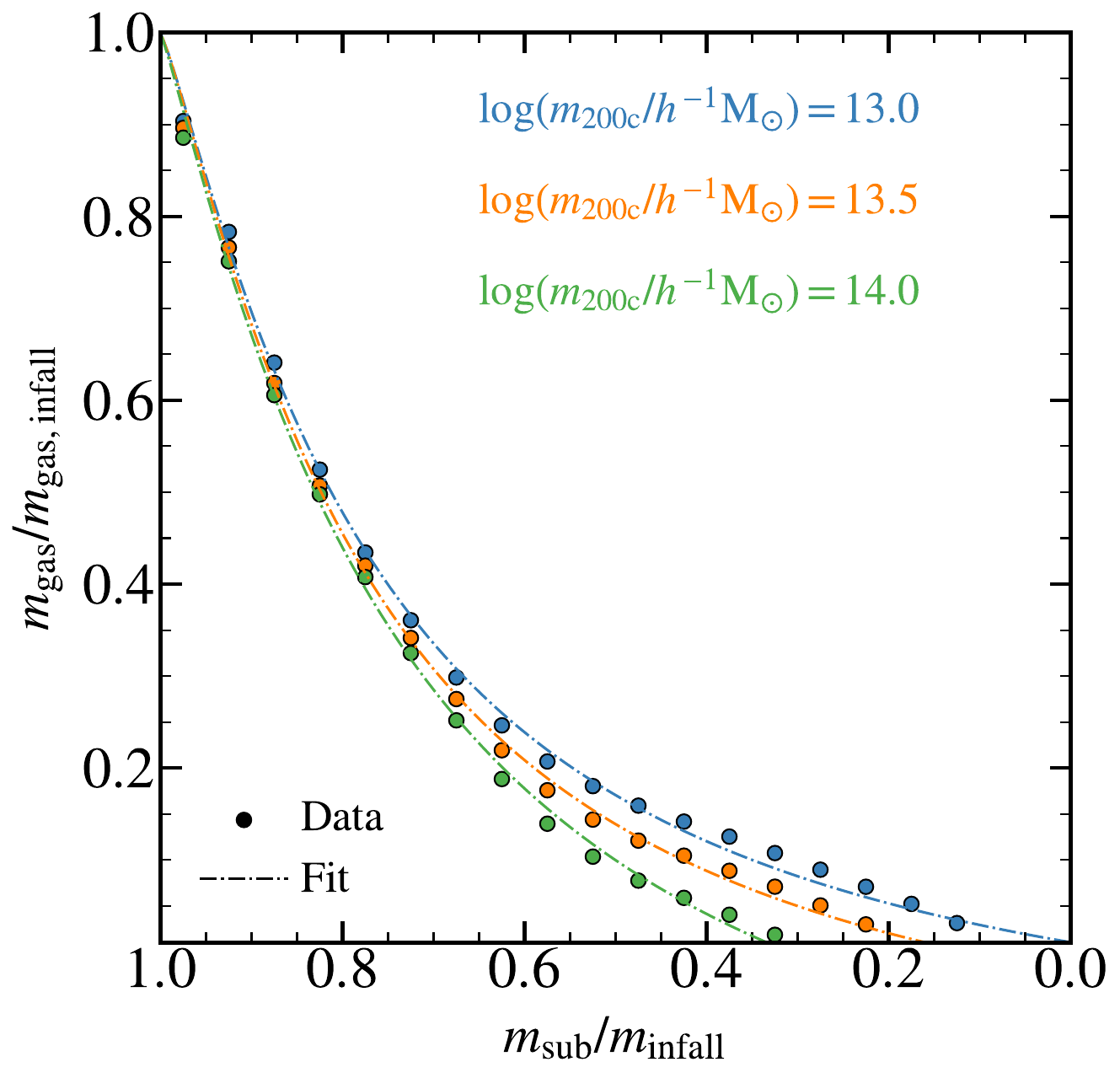}
    \end{subfigure}

    \vspace{0.3cm}

    \begin{subfigure}{0.40\textwidth}
        \centering
        \includegraphics[width=\linewidth]{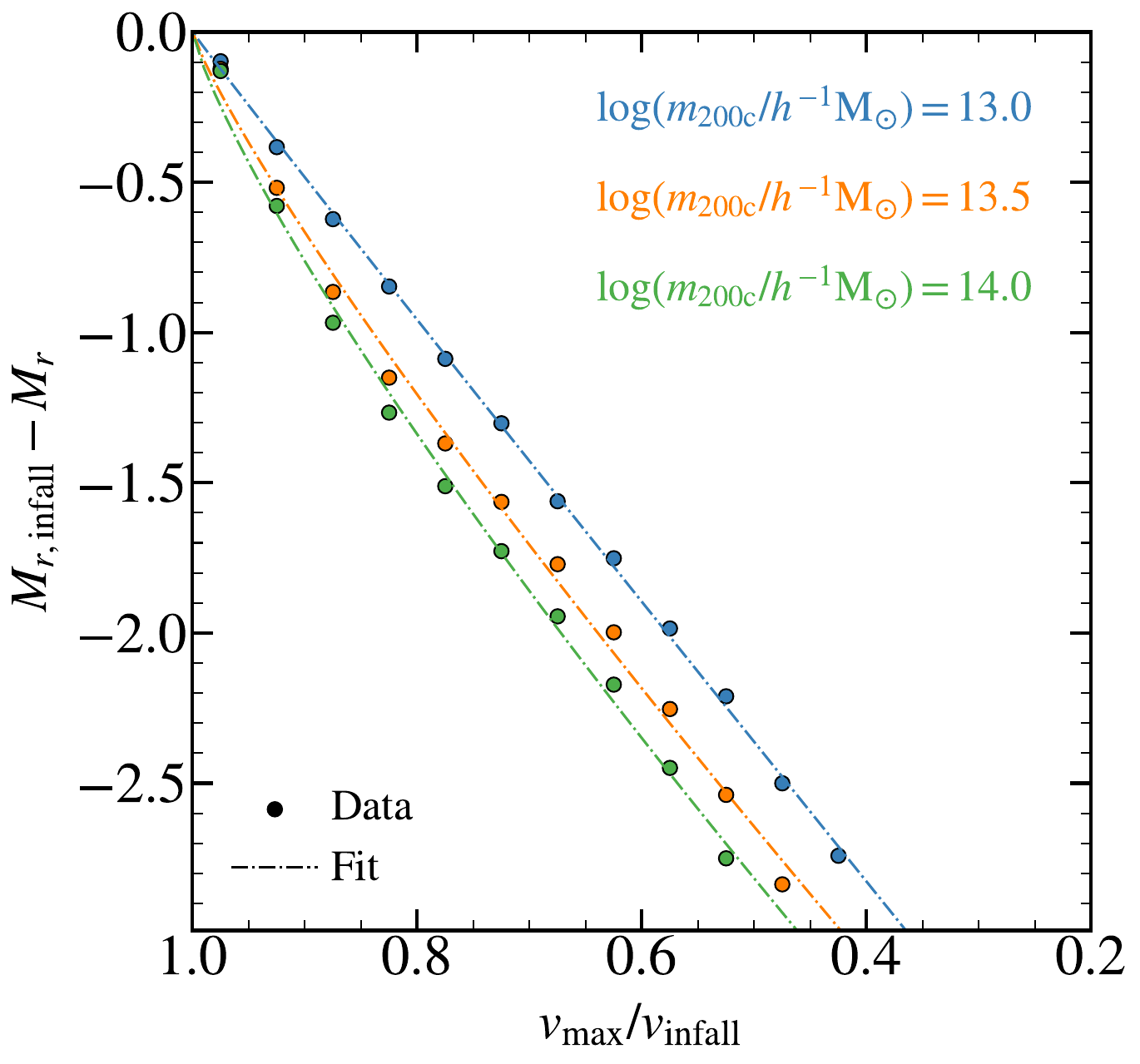}
    \end{subfigure}
    \hspace{0.03\textwidth}
    \begin{subfigure}{0.40\textwidth}
        \centering
        \includegraphics[width=\linewidth]{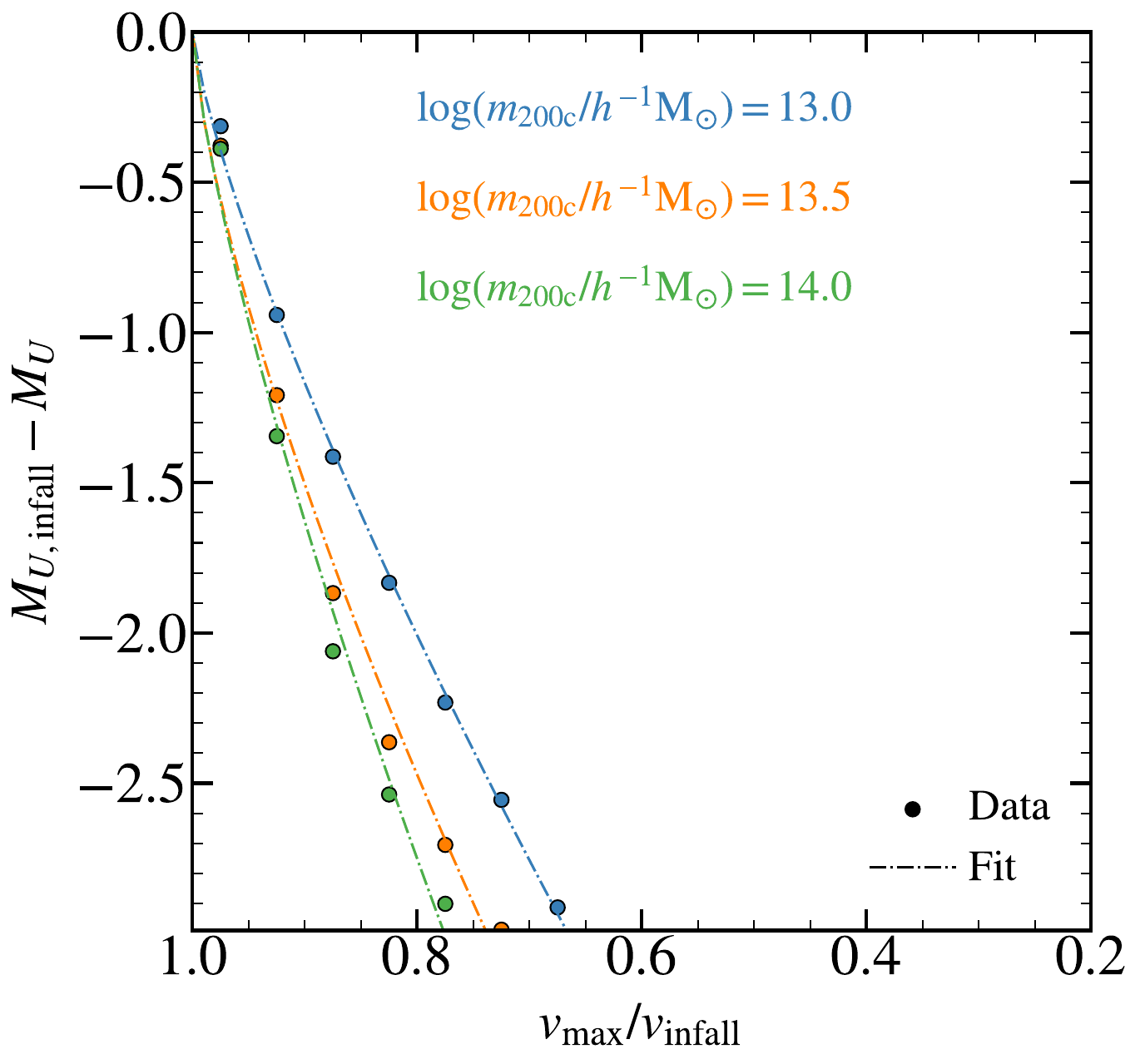}
    \end{subfigure}

    \caption{The evolution of the stellar mass as a function of $\vmax/\vinfall$ (top left panel), the gas mass as a function of $\msub/\minfall$ (top right panel), and the $r$- and $U$-band magnitudes as a function of $\vmax/\vinfall$ (bottom left and right panels) for galaxies with $11.5 < \log(\mpeak/\hMsun) < 12.0$. The symbols represent the data from the \MTNG\ simulation (as shown in Fig.~\ref{Fig:gen_ev}), while the dash-dotted lines represent the analytic models described in Section~\ref{sec:EvBar_fit}. The different colours represent different host halo masses, as labelled.}
    \label{fig:fits}
\end{figure*}

Most of the relations presented in the previous section can be used to some extent to model the evolution of satellite galaxies. Although some trends show lower scatter, weaker dependence on the galaxy sample or host halo mass, and are easier to model, no single proxy provides an optimal description of all galaxy properties. For this reason, we present the full set of relevant relations identified in this work, allowing the choice of proxy to depend on the specific scientific goal. To facilitate the modelling of these properties, we provide a parameterisation for each galaxy property that can be used to describe satellite evolution.

These parameterisations can be combined with known relations between galaxy and halo properties for central galaxies, providing an initial value for each galaxy property. Using these initial values, the subsequent evolution can be modelled as satellites fall into their host haloes. For instance, subhalo abundance matching has been shown to perform well when associating halo properties with galaxy properties, particularly for quantities such as stellar mass or $\Mr$. In the case of $\Mu$, a more complex abundance-matching prescription may be required, such as that proposed by \cite{Ortega:2024,Ortega:2025}. Appendix~\ref{sec:cen} provides a parameterisation for the gas mass that can be used to assign an initial value to central galaxies by characterising their gas mass as a function of $\vmax$.

The key relations we parameterise are the evolution of gas mass as a function of the remaining subhalo mass fraction, and the evolution of stellar mass and magnitudes as a function of the remaining maximum circular velocity fraction. These relations, together with their corresponding parameterisations, are shown in Fig.~\ref{fig:fits}. For simplicity, we show only the results for the intermediate $\mpeak$ sample ($11.5 < \log(\mpeak/\hMsun) < 12.0$). Other galaxy samples show trends similar to those shown here. The decrease in the gas mass is modelled as a shifted and rescaled rational function:

\begin{equation}
\dfrac{{m_{\rm gas}}}{{m_{\rm gas,\ infall}}} = \max\left(0,\dfrac{x}{x+A(1-x)^\beta}(1+\gamma)-\gamma \right),
\end{equation}

\noindent where $x \equiv\msub / \minfall$, and $A$, $\beta$, and $\gamma$ are free parameters. A simpler parameterisation, $x/[x+A(1-x)^\beta]$, is also able to reproduce this trend, although with some limitations at gas fractions below 0.2. As a reference, the best-fitting parameters for the $\log(\mvir/\hMsun) = 13.5$ sample are $A = 6.0 \pm 0.4$, $\beta = 1.17 \pm 0.03$, and $\gamma = 0.021 \pm 0.006$.

{We emphasise, however, that this parameterisation is calibrated to the gas masses assigned to satellite subhaloes in MTNG. As discussed in Appendix~\ref{sec:SUBFIND-HBT}, its absolute normalisation should be interpreted with caution, although comparison with TNG300 suggests that the overall shape of the relation remains reasonably robust.}

For the stellar mass, we model its evolution as a power law with an exponential decay:

\begin{equation}
\dfrac{{m_{\rm stell}}}{{m_{\rm stell,\ infall}}} = \left(\dfrac{A}{A+1-x}\right)e^{-B(1-x)},
\end{equation}

\noindent where $x \equiv \vmax / \vinfall$, and $A$ and $B$ are free parameters. We tested other, more complex parameterisations, but they did not yield significant improvements unless the number of free parameters was increased substantially. As a reference, the best-fitting parameters for the $\log(\mvir/\hMsun) = 13.5$ sample are $A = 16.7 \pm 5.1$ and $B = 1.22 \pm 0.19$. Other parameterisations available in the literature (e.g. \citealt{Smith:2016,He:2026}) focus primarily on the decay of stellar mass, typically modelled as an exponential decline, and do not capture the increase in stellar mass that we find after infall.

Finally, we model the change in magnitude as a linear decay with a negative power law:

\begin{equation}
\Delta M = {\rm M-M_{\rm infall}} = -A(1-x)^B
\end{equation}

\noindent where $x \equiv \vmax / \vinfall$, and $A$ and $B$ are free parameters. We also tested an even simpler form, a linear decay $\Delta M = -A(1-x)$, which provides a reasonably good description over the full $\vmax$ range. As a reference, the best-fitting parameters for the $\log(\mvir/\hMsun) = 13.5$ sample are $A = 4.8 \pm 0.1$ and $B = 0.85 \pm 0.02$ for the $r$ band, and $A = 7.8 \pm 0.8$ and $B = 0.71 \pm 0.06$ for the $U$ band. All fits assume a constant uncertainty for each data point, scaled such that the best fit yields a reduced chi-squared equal to one.

The choice of the best subhalo parameter for modelling these galaxy properties was based on the overall performance and ease of use described in the previous section. Nevertheless, the most suitable subhalo property for characterising a given galaxy property may depend on the specific scientific application. In Appendix~\ref{sec:add_param}, we examine the performance of the parameterisations described in this section when applied to other subhalo properties, finding that they can reproduce the evolution of galaxy properties in most cases. We also provide an additional parameterisation valid across all halo masses, and show that these relations remain valid at $z=1$. These results should facilitate the implementation of analytic models to follow satellite evolution in gravity-only simulations across a broader range of scientific applications.

A more detailed analysis of the dependence of these trends on peak subhalo mass and host halo mass, and of the dependence of higher-order parameters on the scatter of these relations, could provide even more accurate predictions, but this lies beyond the scope of the present paper.
 
\section{The evolution of the mass profile of satellite galaxies}
\label{sec:mass_profile}

\begin{figure*}
\includegraphics[width=0.95\textwidth]{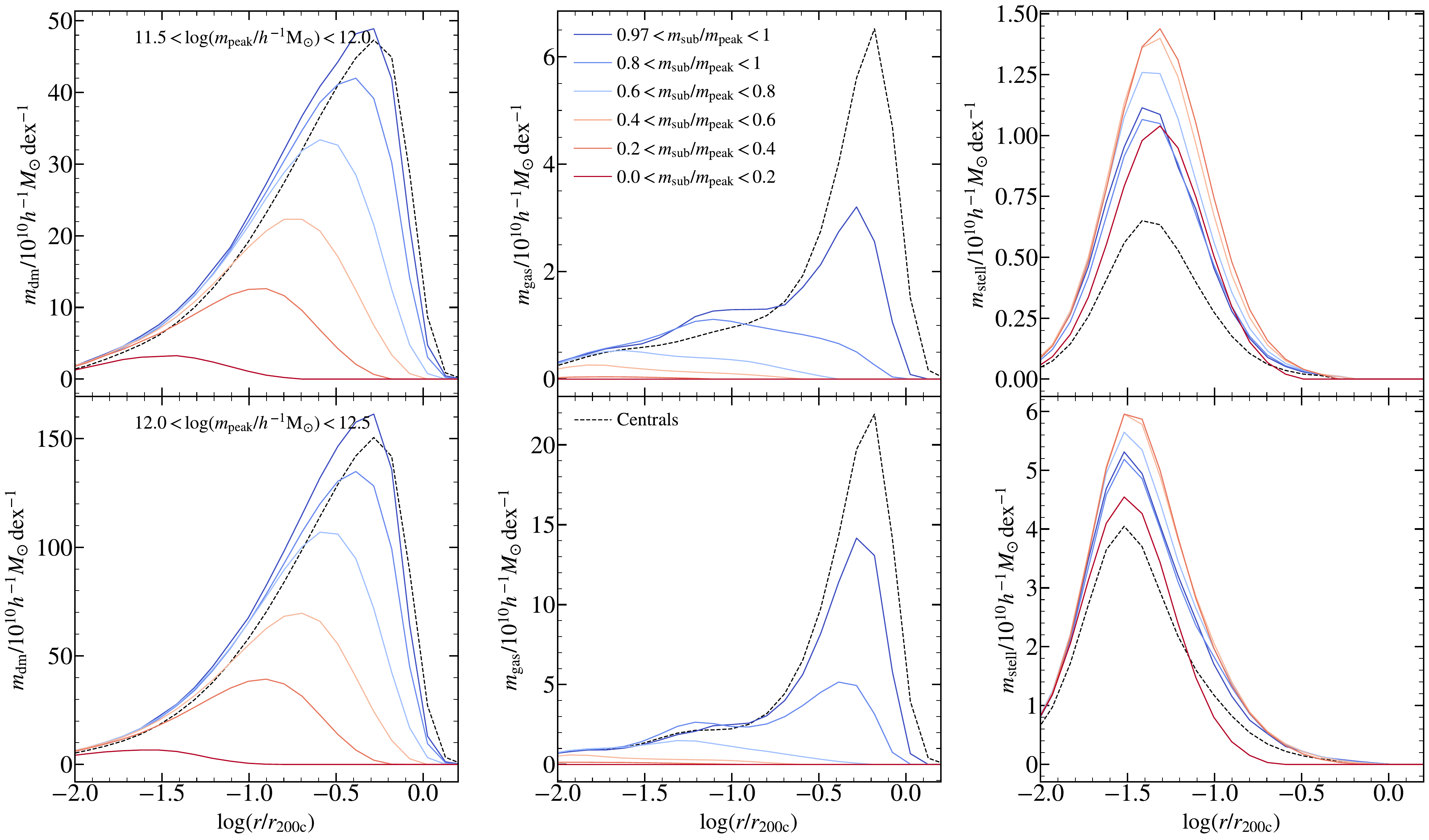}
\caption{The mass profiles of satellite subhaloes with peak subhalo masses between $10^{11.5}$ and $10^{12}\ \hMsun$ (top row) and between $10^{12}$ and $10^{12.5}\ \hMsun$ (bottom row). The first, second, and third columns show the subhalo mass, gas mass, and stellar mass profiles, respectively. The solid coloured lines represent different levels of subhalo mass loss, while the dashed black lines represent the profiles of central subhaloes. Each profile has been normalised by the comoving virial radius ($\rvir$) of the satellite at the moment of infall.}
\label{Fig:profile}
\end{figure*}

\begin{figure*}
\includegraphics[width=0.95\textwidth]{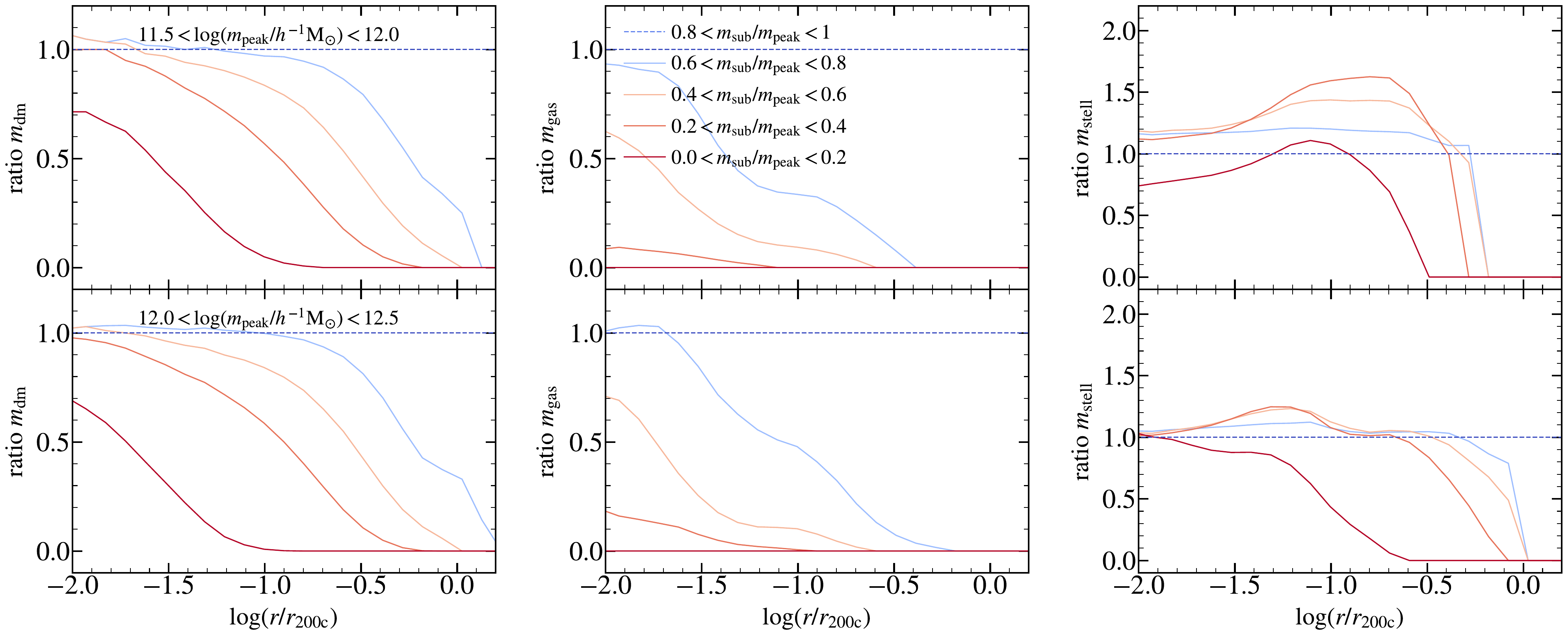}
\caption{Similar to Fig.~\ref{Fig:profile}, but showing the ratio of the profiles to the $0.8 < \msub/\mpeak < 1$ sample.}
\label{Fig:profile_ratio}
\end{figure*}

In the previous section, we quantified and modelled the evolution of satellite galaxy masses and magnitudes as they fall into their host haloes. We now turn to the evolution of the mass profiles of their different components. Since the binding energy is highest in the central regions of galaxies, mass loss is expected to begin in the outer parts of their subhaloes, primarily removing dark matter and gas, and only affecting the stellar component once the galaxies have already lost most of their mass (e.g. \citealt{Cortese:2021}). This picture is consistent with the results in Section~\ref{sec:EvBar}, with the exception that galaxies lose most, if not all, of their gas much faster than they lose their dark matter. We expect the diffuse outer gas to be stripped more efficiently than the dark matter because, although both components are subject to tidal stripping \citep{Tormen:1998, vandenBosch:2018}, only the gas is affected by ram pressure, which can efficiently remove weakly bound material once the external pressure exceeds the local gravitational restoring force (\citealt{McCarthy:2008}, see also \citealt{Ayromlou:2021}).

In Fig.~\ref{Fig:profile}, we show the evolution of the mass profiles of satellites with $11.5 < \log(\mpeak/\hMsun) < 12.0$ and $12.0 < \log(\mpeak/\hMsun) < 12.5$ as a function of distance from the galaxy centre, for the dark matter, gas, and stellar mass components at $z=0$. We express the distance from the galaxy centre in units of the comoving virial radius of the satellites at the moment of infall, which facilitates stacking galaxies that fell in at different times. The stacking is performed independently of host halo mass, since we do not find significant differences in the mass profiles across halo masses. Although more massive haloes provide a more hostile environment, selecting satellites by their $\msub/\mpeak$ ratio allows us to compare galaxies at a similar evolutionary stage, largely independently of host halo mass. We do not include the lowest-$\mpeak$ galaxy sample because the number of particles per object is low, especially for the stellar mass component. The y-axis represents the total mass in a spherical shell (i.e. we do not normalise by the shell volume). The different coloured lines represent the $\msub/\mpeak$ bins. This ratio is similar to $\msub/\minfall$ and allows us to characterise the evolution of the profiles independently of the definitions of central and satellite galaxies. We group galaxies in bins of 0.2 in $\msub/\mpeak$ (as labelled). We also include a galaxy sample with $\msub/\mpeak > 0.97$, which represents satellites shortly after infall, as well as the mass profile of central galaxies.

In agreement with the results in Section~\ref{sec:EvBar}, we find that the gas component is the first to be removed, initially from the outer regions and later from the inner ones. The dark matter component shows a similar behaviour, but on a slower timescale. The stellar mass component first increases and only later begins to decrease, and it is not yet clear whether this stripping proceeds from the outside in or follows a more complex behaviour. These trends are shown more clearly in Fig.~\ref{Fig:profile_ratio}, where we present the ratio of the different subhalo-mass-ratio samples to the $0.8 < \msub/\mpeak < 1$ sample. We do not use the sample of satellites shortly after infall for the denominator, since it is noisier because of its limited number of galaxies. It is important to note that, although most of the mass is lost from the outer regions, there is a continuous loss of gas in the central regions. This suggests that the simple picture in which mass is removed smoothly from the outside in provides a useful approximation in many cases, but does not fully capture the complexity of satellite evolution. The figure may also indicate that dark matter is stripped from the central regions, although this could partly reflect the fact that the dark matter mass profiles do not fully converge to unity on the scales shown here (see \citealt{Stucker:2023} for a discussion of this issue).

When comparing the evolution of the two satellite samples selected by $\mpeak$, we find overall qualitative agreement in how galaxies lose dark matter and gas, but not in how their stellar mass component evolves. Lower-mass galaxies show a more pronounced increase in stellar mass, peaking at around 10\% of their virial radius. Only when the subhalo mass has fallen significantly does the stellar mass begin to decrease. The more massive sample loses mass first from the outer regions and later from the inner ones, while the less massive sample shows no clear trend. The overall good agreement between the results for the two galaxy samples, except for the stellar mass component, which contributes only a small fraction of the total subhalo mass, suggests that the way satellite galaxies lose mass may be largely independent of the masses of the satellite and the host halo. Unfortunately, given the limited number of samples that can be studied because of the resolution and volume limitations, and the fact that we can examine only one galaxy formation prescription, we cannot generalise this result to a universal relation. A more detailed study of satellite evolution in other simulations is left for future work.

In addition to the evolution of satellite galaxies, Fig.~\ref{Fig:profile} also reveals differences between the profiles of satellites shortly after infall and central galaxies. These results are independent of the normalisation adopted for the distance to the galaxy centre, as well as of the bin width in $\msub/\mpeak$. These differences arise because central galaxies are intrinsically different from satellites, even shortly after infall, with centrals having more gas and less stellar mass, consistent with the results of \cite{Contreras:2015} for semi-analytical models and \cite{He:2026} for hydrodynamic simulations. This may be partly related to the fact that, at $z=0$, more concentrated haloes tend to reside in denser environments, an effect known as halo assembly bias (e.g. \citealt{Gao:2005, Gao:2007}), and these environments are also known to host more massive and less star-forming galaxies (e.g. \citealt{Zehavi:2018, Artale:2018}). By selecting massive satellite galaxies, as done in this work, we are selecting objects in denser environments than those associated with a less-biased sample of central galaxies. The differences between centrals and satellites are particularly pronounced for the gas component. Part of this difference also arises because satellites reach their peak gas mass 0.5--1 Gyr before they reach their peak subhalo mass, and up to 2 Gyr before they become satellites. This suggests that the infall process, and therefore gas stripping, begins earlier than implied by the moment at which the galaxy is formally classified as a satellite, while the subhalo may still be increasing in mass, probably owing to interactions with the outer regions of the host halo it will eventually merge into. Since gas loss is abrupt, with galaxies losing most of their gas within 2--3 Gyr after infall, satellites may already have lost a considerable fraction of their gas before they are classified as satellites or before they reach their peak subhalo mass, which may explain the behaviour seen in Figs.~\ref{Fig:profile} and~\ref{Fig:profile_ratio}. This result suggests that, for some applications, the usual central--satellite classification may not be ideal. Empirical models such as HODs rely entirely on this classification and generally do not distinguish between the overall central population and centrals that are about to become satellites. Our results therefore suggest that a more conservative definition of satellite galaxies, in which systems are classified as satellites at earlier times, may be required. Although the focus of this paper is not the intrinsic differences between central and satellite galaxies, Appendix~\ref{sec:cen_infall_sat} includes a more detailed analysis of the origin of the differences discussed in this section.
 
\section{Summary}
\label{sec:summary}

In this work, we examine the evolution of the masses and mass profiles of satellite galaxies in the cosmological hydrodynamic \MTNG\ simulation. We first identify satellite galaxies at $z=0$ and trace them back to $z\sim 13$. We quantify the evolution of their dark matter mass, stellar mass, gas mass, and their $r$- and $U$-band magnitudes as a function of the ratio of the current subhalo mass to its value at infall ($\msub/\minfall$), the pericentric distance, the time since infall, and the ratio of the current maximum circular velocity to its value at infall ($\vmax/\vinfall$) (Fig.~\ref{Fig:gen_ev}). All of these quantities can be obtained from gravity-only simulations with merger trees and can therefore be used to model the evolution of galaxy properties in these simpler simulations. The main findings of this part of the work are as follows:

\begin{itemize}
    
\item The stellar mass increases as the satellite evolves, and only begins to decrease once the galaxy has lost a substantial fraction of its subhalo mass or maximum circular velocity. We find that the remaining subhalo mass fraction and the maximum circular velocity ratio can both be used to characterise stellar mass evolution.

\item The gas mass decreases abruptly as the satellite evolves, with most of the gas mass lost once the galaxy has lost only about half of its total subhalo mass. The remaining subhalo mass fraction and the time since infall provide the best characterisation of this evolution, with the former showing lower scatter. {As discussed in Appendix~\ref{sec:SUBFIND-HBT}, the absolute gas-mass scale should nevertheless be treated with caution, as the gas masses assigned to satellite subhaloes in MTNG may be biased low.}

\item The magnitudes in both bands are reasonably well described by all the tested proxies. Both bands follow similar trends, although the $U$-band magnitude fades faster than the $r$-band magnitude.

\item The trends found at $z=0$ are largely consistent with those found at $z=1$, suggesting that they may be broadly universal and therefore applicable to models of satellite evolution across different infall redshifts (Fig.~\ref{Fig:gen_ev_z1}).
\end{itemize}

Most trends show little dependence on the peak subhalo mass of the galaxies or on the mass of their host haloes. The $\vmax$ ratio is found to be particularly insensitive to these selections in most cases, with the exception of the gas, for which the remaining subhalo mass fraction appears to be the least sensitive proxy. These results suggest that simple parametric models can successfully characterise satellite evolution.

We provide simple parameterisations for the evolution of the baryonic properties studied in this work, with the aim of modelling satellites in empirical, semi-empirical, and semi-analytical models. We characterise the evolution of stellar mass and magnitudes as a function of $\vmax/\vinfall$, and the evolution of gas mass as a function of $\msub/\minfall$. We find that all these relations can be reproduced reasonably well with only two or three free parameters (Fig.~\ref{fig:fits}). These parameterisations can be combined with a simple subhalo abundance-matching prescription to assign initial galaxy properties before satellite infall, thus enabling the assignment of galaxy properties to subhaloes in gravity-only simulations. To complement this framework, Appendix~\ref{sec:cen} provides a parameterisation to assign gas mass to central subhaloes in gravity-only simulations, since this property is not well characterised by the standard subhalo abundance-matching formalism. We note, however, that because there are intrinsic differences between standard centrals and centrals shortly before becoming satellites, or satellites shortly after infall (Appendix~\ref{sec:cen_infall_sat}), this step may require additional adjustments to perform optimally.

We then focus on the evolution of the dark matter, stellar mass, and gas mass profiles of satellites at different evolutionary stages. We select galaxies at $z=0$ with different ratios of the current subhalo mass to the maximum subhalo mass reached during their evolutionary history ($\msub/\mpeak$), for two galaxy samples defined by their $\mpeak$ values (Figs.~\ref{Fig:profile} and~\ref{Fig:profile_ratio}). Since we find similar trends across different host halo masses, we stack all galaxies together to maximise the statistical power of the results. The main findings of this part of the work are as follows:

\begin{itemize}
\item In agreement with the results shown in Fig.~\ref{Fig:gen_ev}, we find that the gas mass is lost much faster than the total subhalo mass, while the stellar mass increases as the satellite evolves and only later begins to decrease, once most of its subhalo mass has been lost.

\item The loss of all mass components begins preferentially in the outer regions of the galaxy, and as the satellite evolves, mass loss extends progressively towards the inner regions.

\item A significant fraction of the gas mass does not follow this simple trend, with a continuous loss of mass occurring at all radii. 

\item While the mass profiles differ between the two $\mpeak$ samples, the ratio of the profiles of satellites in a late evolutionary stage to those of satellites shortly after infall is similar for the gas and dark matter components, which together account for most of the subhalo mass.

\item The stellar-mass profiles show clearer differences between the two $\mpeak$ samples, mostly owing to the different starburst episodes experienced by satellites after infall.
\end{itemize}

The overall agreement in the evolution of the gas and dark matter components suggests, once again, that the internal mass profiles of satellite galaxies could be parametrised relatively easily if needed. That said, the limited number of galaxy samples studied in this work, together with the use of a single galaxy formation model, strongly limits our ability to make stronger claims about the universality of satellite evolution. Nevertheless, the trends identified here suggest that these relations could be recalibrated for different galaxy formation models and used to improve the modelling of satellite galaxies.

In \cite{Contreras:2026a}, we studied how baryons alter the orbits of satellite galaxies, and quantified and parameterised the changes in the positions and velocities of galaxies relative to the subhaloes of gravity-only simulations. The results of the present work extend those findings by providing a more complete picture of how galaxies evolve after becoming satellites. In future work, we plan to incorporate the results of both studies into the extended Subhalo Abundance Matching model (SHAMe; \citealt{Contreras:2021_SHAMgab, Contreras:2021_SHAMe}) to improve clustering predictions in the non-linear regime. In addition, the differences in the mass profiles will be used in the \texttt{BACCO} baryonification model \citep{Arico:2020, Arico:2021, Arico:2024, Burger2025} to further improve the precision of baryonification emulators. These results will also be used to extend to smaller scales the range of validity of galaxy--galaxy lensing modeling, like the emulator \texttt{GalaxyEmu} \citep{Contreras:2023_MTNG, Mahony:2026} and the model by \cite{Zennaro:2025}, as well as to improve the prescriptions for the satellite contribution to the kinetic Sunyaev Zel'dovich signal \citep[see e.g.][]{McCarthy:2025}.

\begin{acknowledgements}
We thank Jonas Chaves-Montero and Jens St\"ucker for some useful suggestions and comments.
SC acknowledges the support of the ``Ram\'on y Cajal'' fellowship (RYC2023-043783-I). SC also acknowledges the support of the ``Ayudas para Atracci\'on de Investigadores con Alto Potencial'' (2025/00000640) from Universidad de Sevilla.
REA received support from grant PID2024-161003NB-I00 funded by MICIU/AEI/10.13039/501100011033 and by ERDF/EU.
SB is supported by the UKRI Future Leaders Fellowship [grant numbers MR/V023381/1 and UKRI2044].

{\it Author Contributions Statement:} The idea for this project was developed by SC, with essential contributions from REA and GA. SC analysed the data, performed the calculations, and wrote the manuscript. REA, GA and LO provided critical comments throughout the development of the project. SB, LH, RP, and VS developed the MTNG simulation and provided feedback on the manuscript.
\end{acknowledgements}
\bibliographystyle{aa}
\bibliography{aa.bib}

\begin{appendix}

\section{Modelling the gas content of central galaxies}
\label{sec:cen}
\begin{figure}
\includegraphics[width=0.45\textwidth]{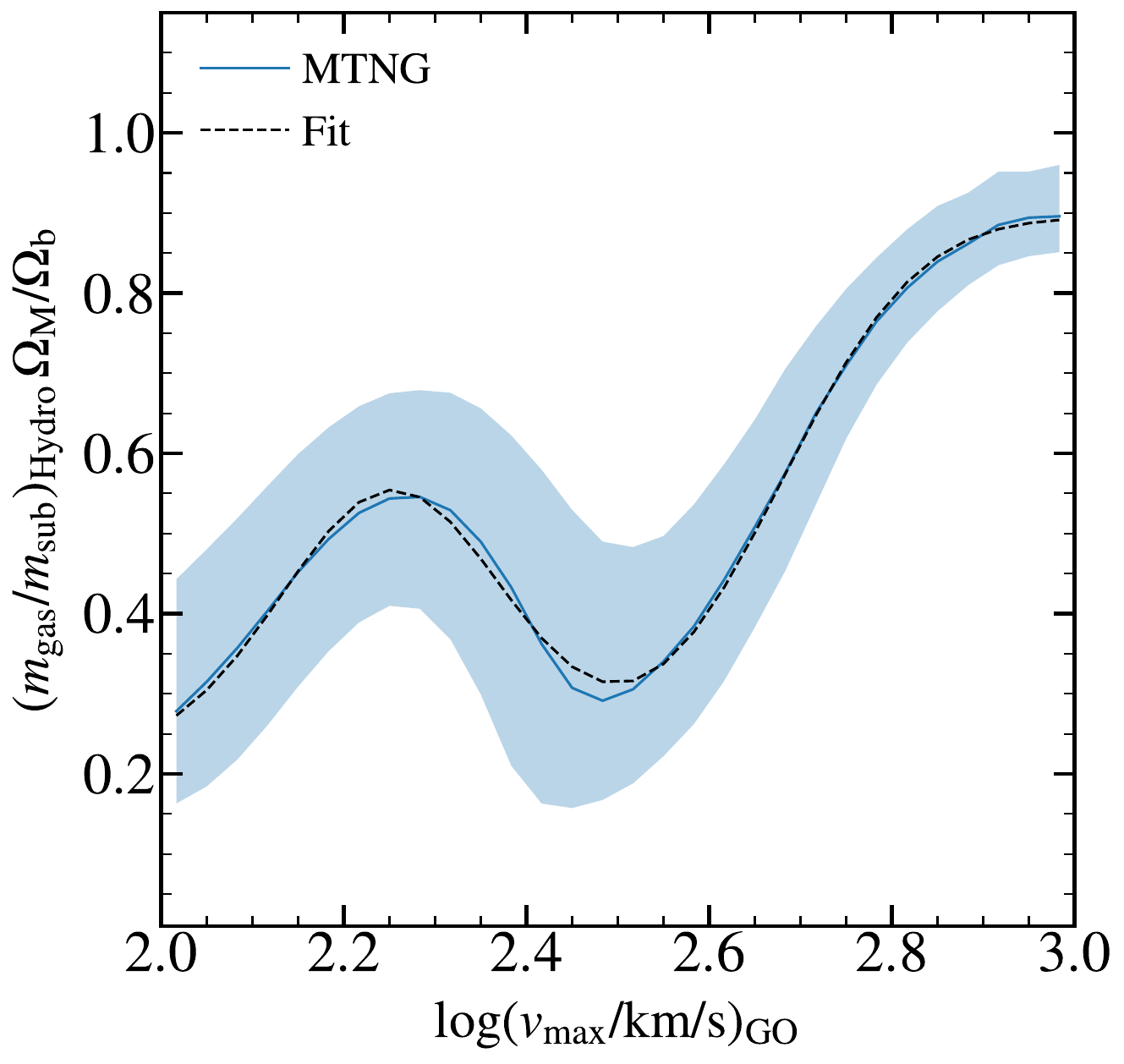}
\caption{The gas fraction of central subhaloes in the \MTNG\ simulation, normalised by $\Omb/\OmM$, as a function of the peak value of the maximum circular velocity reached by each galaxy during its evolutionary history (i.e. $\vpeak$). The solid coloured line represents the median of the distribution, while the shaded region represents the 68th percentile. The dashed black line represents the fit to this distribution (see Appendix~\ref{sec:cen} for more details).}
\label{Fig:cen_fit}
\end{figure}

In Section~\ref{sec:EvBar_fit}, we present a parameterisation for the evolution of satellite properties as a function of subhalo properties available in gravity-only simulations. These parameterisations can be used by empirical, semi-empirical, and semi-analytical models to improve their description of satellite galaxies. These relations are expressed as the ratios of stellar mass and gas mass to their values at infall, and as changes in the magnitudes. Modelling these evolutions therefore requires an initial value for these properties at the time when the galaxies were still centrals. For stellar mass and the magnitudes, a subhalo abundance-matching formalism can be used, in which a halo property such as halo mass or $\vmax$ is rank-ordered and matched to the expected stellar mass or magnitude distribution. Subhalo properties such as peak subhalo mass and peak maximum circular velocity can also be used, although for central galaxies they may not yield a significant improvement in performance. In the case of the $U$-band magnitude, a more complex abundance-matching prescription, such as that proposed by \cite{Ortega:2024, Ortega:2025}, may be required to model this property accurately.

The only property that cannot be fully modelled with a standard abundance-matching formalism is the total gas mass. We expect the total gas mass, normalised by the subhalo mass, to be close to $\Omb/\OmM$, which corresponds to the maximum baryonic mass available within the halo. The difference between this approximation and the true value arises from the stellar-mass and black-hole-mass components, which are expected to be subdominant, and from gas that has been expelled or redistributed by feedback processes. We express this ratio as a function of $\vmax$ rather than subhalo mass, since we find that the corresponding relation shows slightly less scatter, although in practice either quantity could be used.

To facilitate the implementation of this relation in galaxy population models, which are usually built on gravity-only simulations, we characterise the gas-to-subhalo-mass ratio as a function of the $\vmax$ of the matched haloes in the \MTNGdmo\ gravity-only simulation. We perform the halo matching between the two simulations by identifying pairs of objects with similar positions and masses. To ensure a clean selection, we use only haloes that are matched bidirectionally, i.e. haloes for which an object in the hydrodynamic simulation is matched to an object in the gravity-only simulation, and for which that same object in the gravity-only simulation is in turn matched back to the original halo in the hydrodynamic simulation.

The gas fraction of central galaxies in the \MTNG\ simulation at $z=0$ is shown in Fig.~\ref{Fig:cen_fit}. This relation shows a local peak at about $\vmax \sim 200\ {\rm km\,s^{-1}}$, which corresponds to a halo mass of $\sim 10^{12}\ \hMsun$, the scale at which AGN feedback becomes efficient. The gas fraction then decreases before rising again, reaching values close to the cosmological baryon fraction. This shape is similar to that found in other hydrodynamic simulations (e.g. \citealt{Schaye:2023}), suggesting that, although the relation may differ across galaxy formation models, the same fitting function could still be used after recalibrating its free parameters. We parameterise this relation as the sum of a Gaussian, which reproduces the first peak, and an error function, which models the subsequent increase:

\begin{equation}
f(x)=c + A\,g(x) + B\,s(x)
\end{equation}

\noindent where $x \equiv \log(\vmax/{\rm km\,s^{-1}})$, $c$ is a constant, and $g(x)$ is a Gaussian:

\begin{equation}
g(x)=\exp\left[-\frac{1}{2}\left(\frac{x-\mu}{\sigma}\right)^2\right]
\end{equation}

\noindent while

\begin{equation}
s(x)=\frac{1}{2}\left[1+\operatorname{erf}\left(\frac{x-x_t}{w}\right)\right]
\end{equation}

\noindent is the error function. This form successfully reproduces the relation between gas fraction and $\vmax$ over the full $\vmax$ range. The best-fitting parameters are $c$,~$A$,~$\mu$,~$\sigma$,~$B$,~$x_t$,~and~$w$ = 0.167, 0.249, 2.254, 0.122, 0.505, 2.677, and 0.168.

As discussed in Section~\ref{sec:mass_profile} and as will be discussed in Appendix~\ref{sec:cen_infall_sat}, there are intrinsic differences between central galaxies and satellites shortly after infall, with the latter population potentially providing a more appropriate starting point for modelling satellite evolution. We therefore also examine the gas-mass ratio for these galaxies and fit Equation~(A.1), finding the best-fitting parameters to be $c$,~$A$,~$\mu$,~$\sigma$,~$B$,~$x_t$,~and~$w$ = 0.140, 0.319, 2.248, 0.122, 0.797, 2.68, and 0.132. To match satellites shortly after infall to the dark matter haloes in the gravity-only simulation, we use the method developed in \cite{Contreras:2026a}, which consists of tracing satellites back to the point at which they were still centrals and had a subhalo mass equal to 80\% of the peak subhalo mass they would eventually reach, and performing the matching at that redshift.

\section{The differences between central galaxies and satellites shortly after infall}
\label{sec:cen_infall_sat}

\begin{figure}
\includegraphics[width=0.45\textwidth]{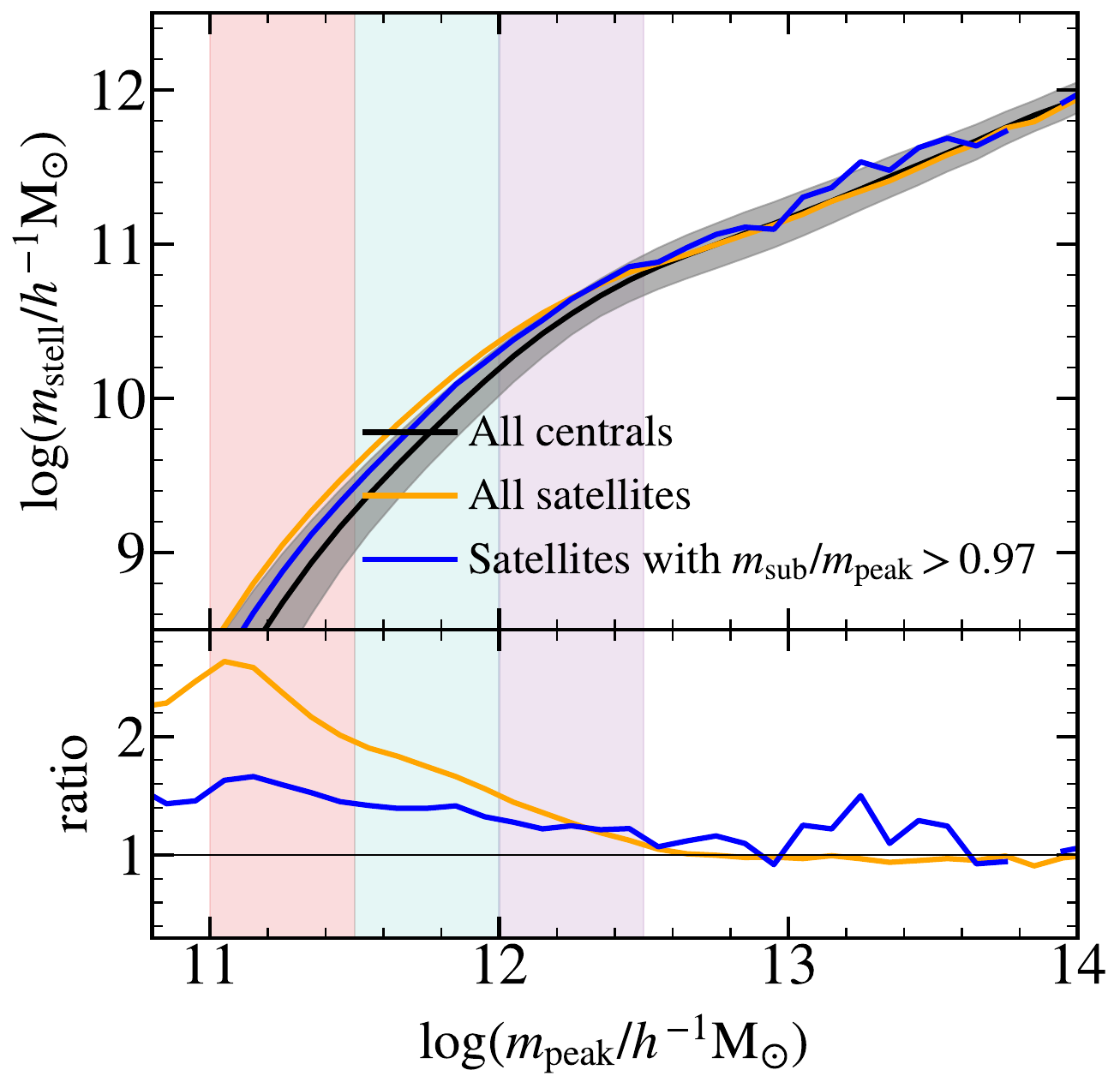}
\includegraphics[width=0.45\textwidth]{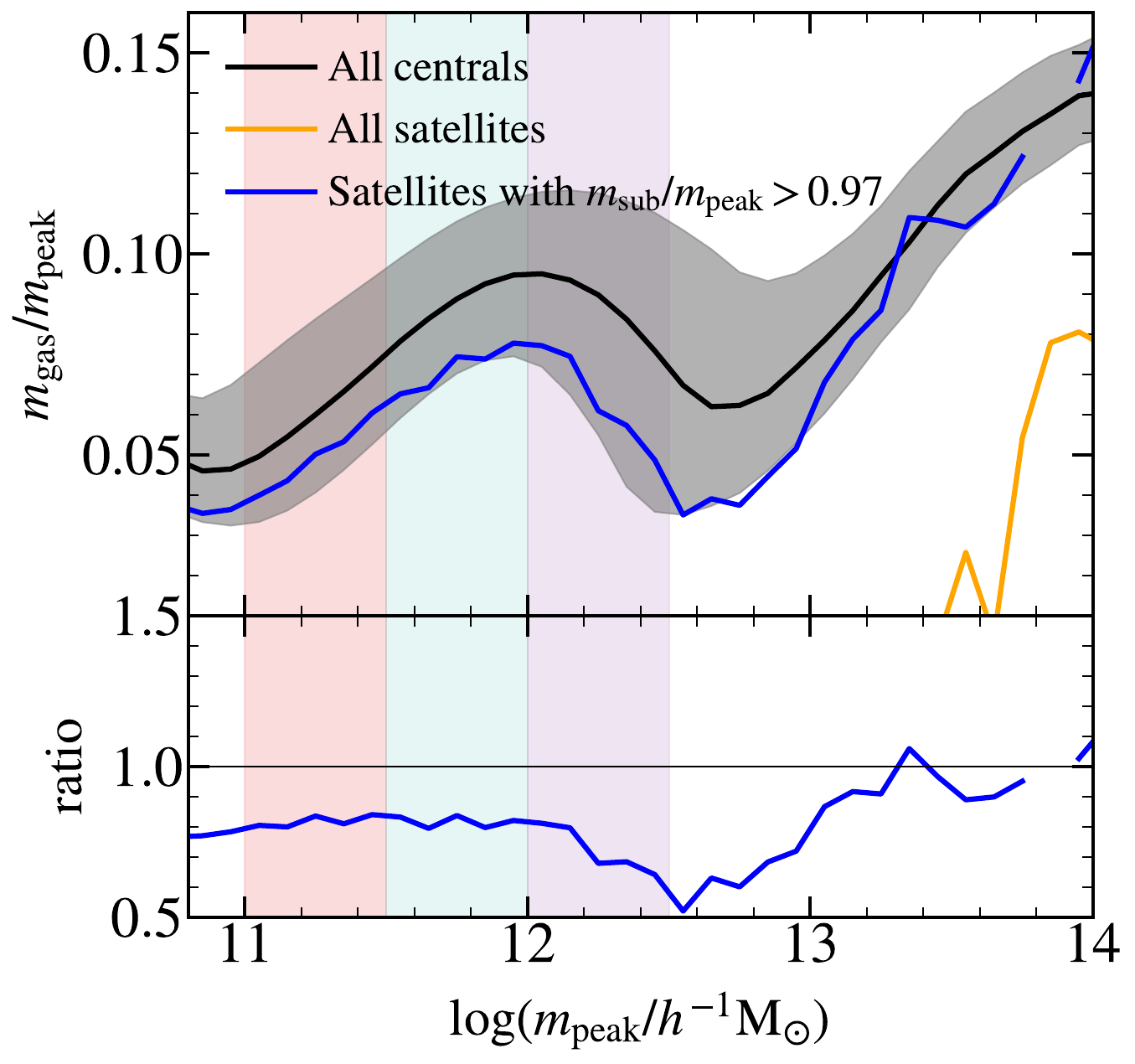}
\caption{(Top) Stellar mass as a function of peak subhalo mass. The black solid line and the shaded regions correspond to the median of the distribution for central galaxies and the 16th and 84th percentiles, respectively. The orange line shows the median stellar mass for all satellite galaxies, while the blue line shows only satellites with $\msub/\mpeak > 0.97$ (representing satellites shortly after infall). (Bottom) Similar to the top panel, but for gas mass normalised by peak subhalo mass. The lower panel in each figure shows the ratio of the coloured lines to the black line (central galaxies).}
\label{Fig:mpeak_rel_censat}
\end{figure}

\begin{figure}
\includegraphics[width=0.45\textwidth]{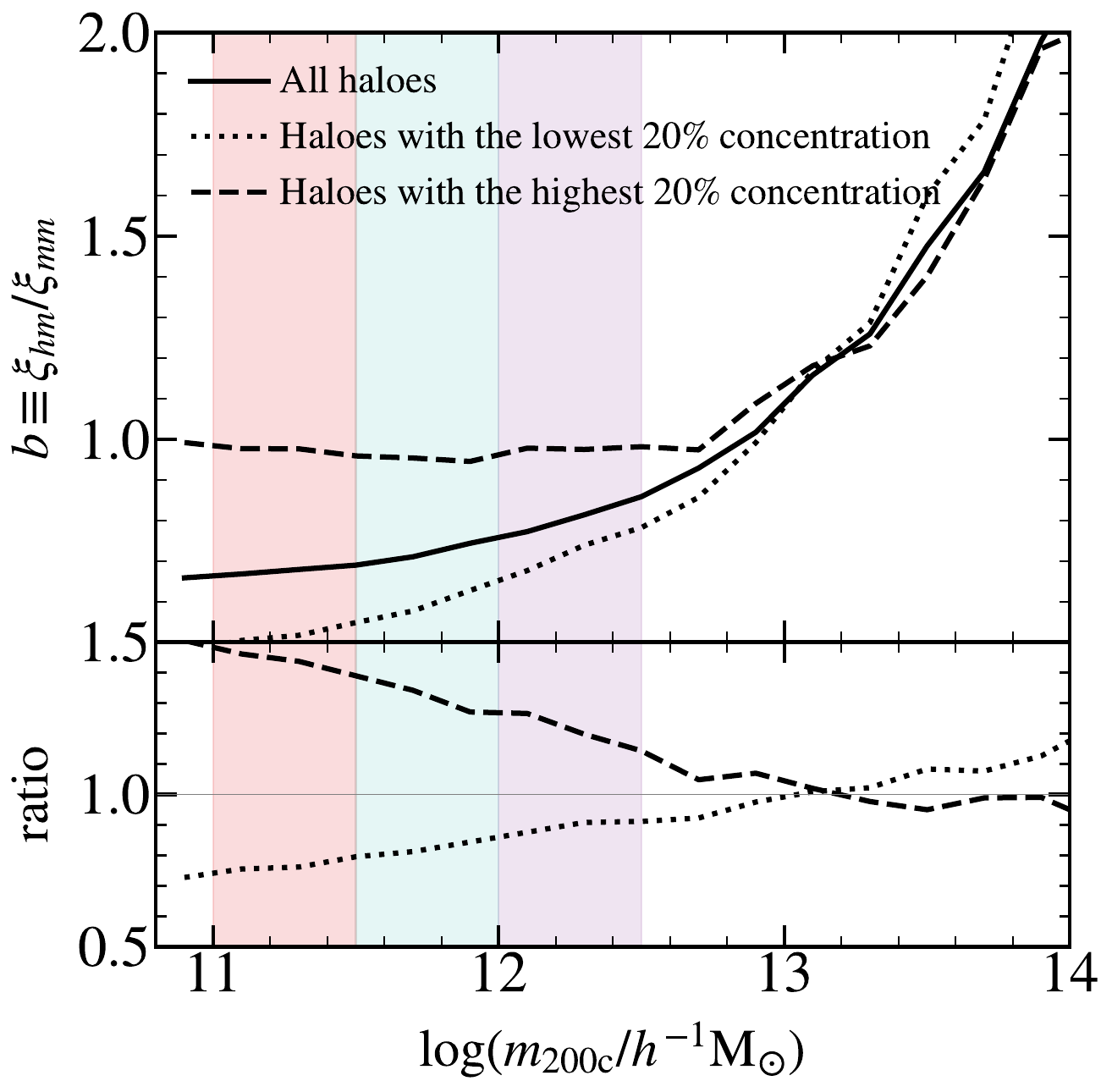}
\caption{The halo bias in the \MTNGdmo\ simulation, computed as the ratio of the halo--matter cross-correlation function to the theoretical matter correlation function. The dotted and dashed lines show the biases of the haloes in the lower and upper 20\% of the concentration distribution, respectively, in bins of halo mass (i.e. the concentration halo assembly bias). The shaded regions correspond to the bins in $\mpeak$ used in this work, which should approximately map onto similar halo-mass ranges.}
\label{Fig:hab}
\end{figure}

\begin{figure}
\includegraphics[width=0.45\textwidth]{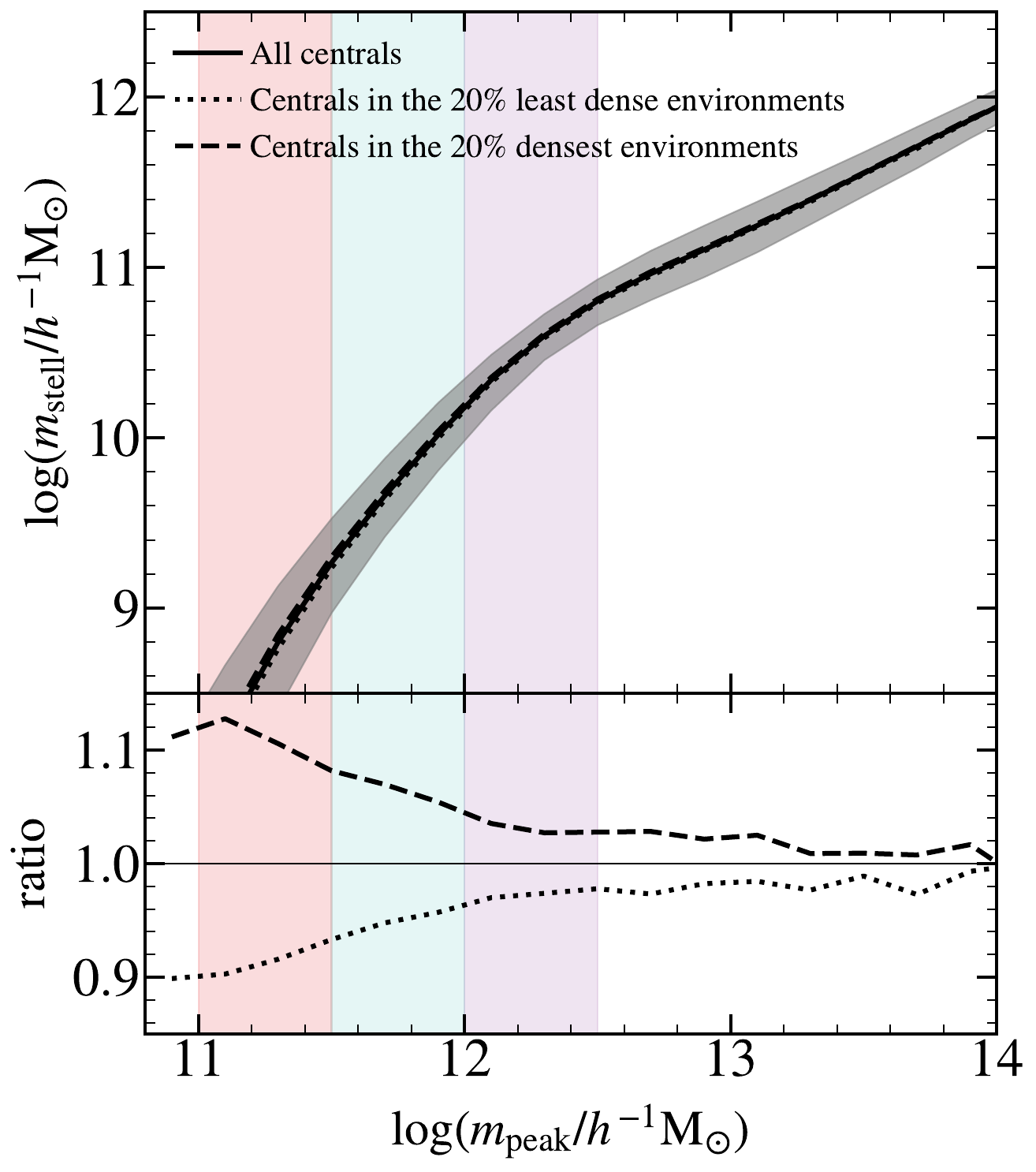}
\includegraphics[width=0.45\textwidth]{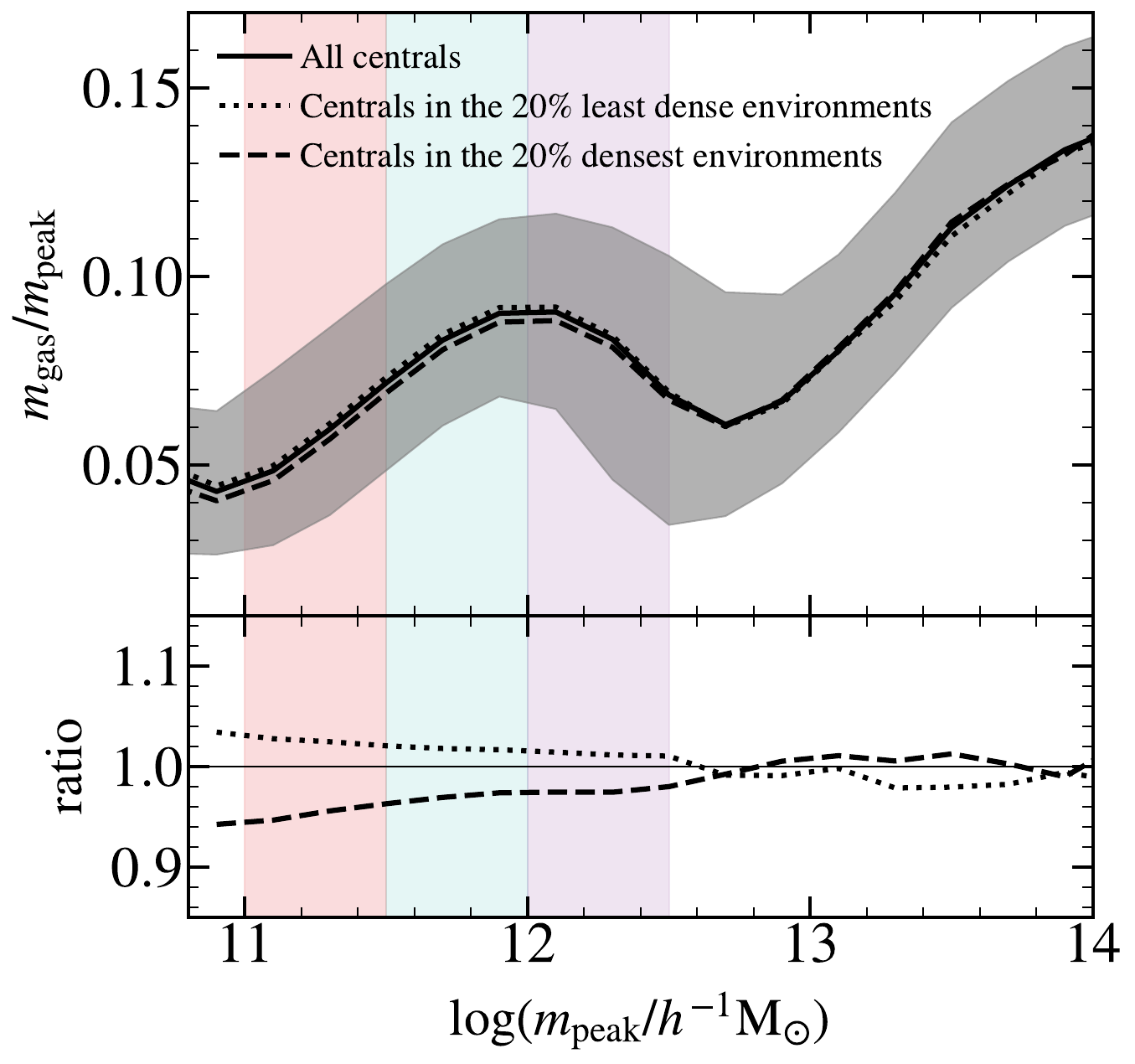}
\caption{(Top) The stellar mass distribution as a function of the peak subhalo mass. The black solid line and the shaded regions correspond to the median of the distribution for central galaxies and the 16th and 84th percentiles, respectively. The dashed and dotted lines represent central galaxies in haloes residing in the 20\% most and least dense environments. The environmental classification is based on the individual bias-per-object and is performed in logarithmic bins of halo mass. (Bottom) Similar to the top panel, but for gas mass normalised by peak subhalo mass.}
\label{Fig:mpeak_rel_env}
\end{figure}

\begin{figure*}
\includegraphics[width=0.95\textwidth]{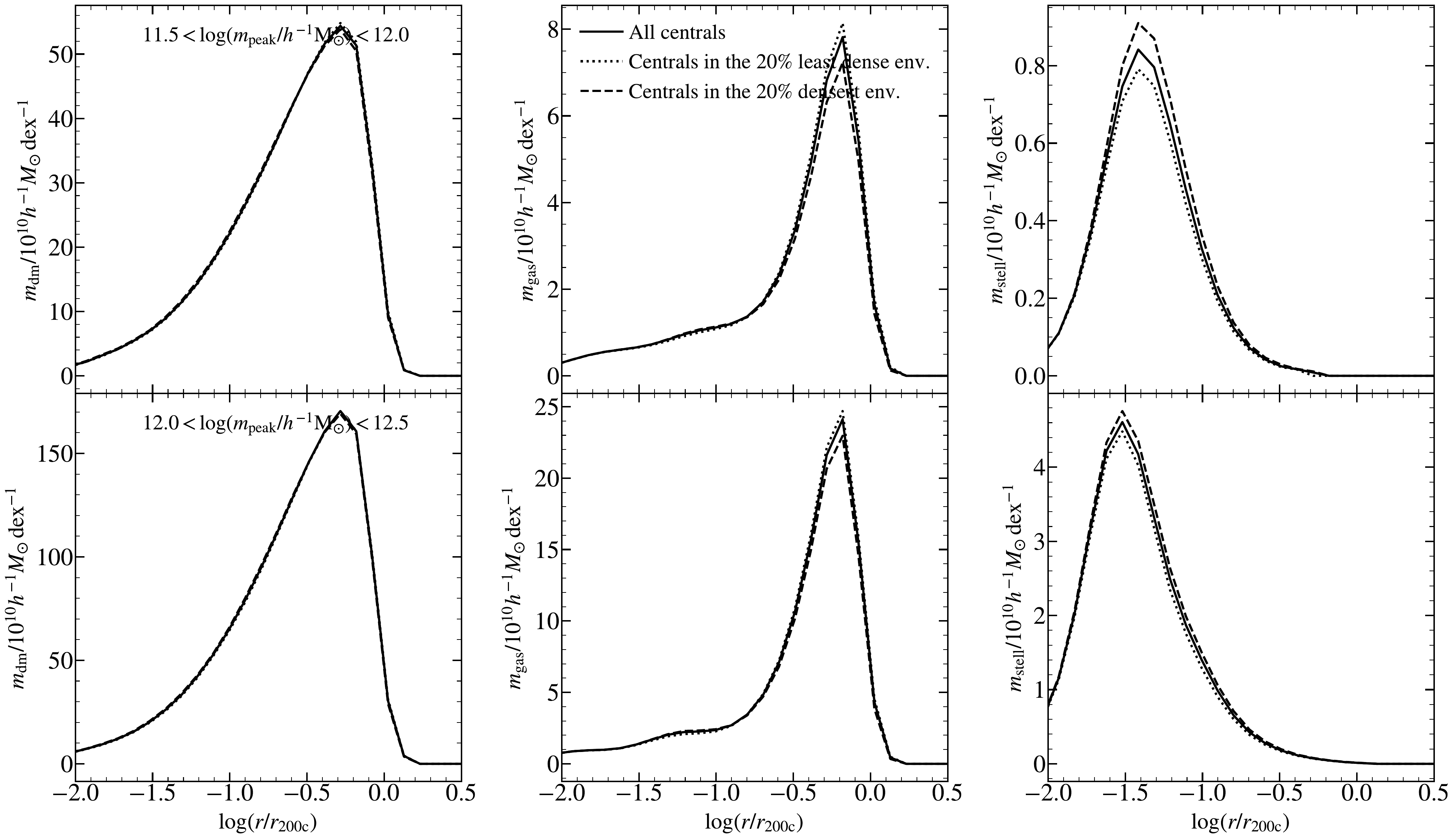}
\caption{Similar to Fig.~\ref{Fig:profile}, but for centrals in the 20\% densest and least dense environments. The environmental classification is based on the individual bias-per-object and is performed in logarithmic bins of halo mass.}
\label{Fig:profile_centrals}
\end{figure*}

\begin{table}
\caption{The difference in look-back time and redshift between the infall time of satellite galaxies and the time at which they reach their peak subhalo mass (i.e. $\mpeak$), peak stellar mass, and peak gas mass. A positive (negative) value indicates that the corresponding peak mass was reached before (after) the galaxy became a satellite.}
\begin{tabular}{c c c c} 

  ${10^{11.5} < \mpeak}/\hMsun < 10^{12}$  & $\msub$  & $\Mstell$ & $\Mgas$ \\ [0.5ex] 
 \hline
 $\rm t_{peak} - t_{infall}$ [Gyr] & 0.68 & -2.91 & 1.81  \\ 
 $\rm z_{peak} - z_{infall}$ & 0.11 & -0.37 & 0.33  \\ 
 \\
   ${10^{12} < \mpeak}/\hMsun < 10^{12.5}\ $  & $\msub$  & $\Mstell$ & $\Mgas$ \\ [0.5ex] 
    \hline
 $\rm t_{peak} - t_{infall}$ [Gyr] & 0.45 & -2.17 & 1.96 \\ 
 $\rm z_{peak} - z_{infall}$ & 0.07 & -0.23 & 0.36\\ 
\end{tabular}
\label{Table:times}
\end{table}

\begin{table}
\caption{The mean change in total subhalo mass ($\msub$), stellar mass ($\Mstell$), and gas mass ($\Mgas$) for central and satellite galaxies around the time of infall. The mass changes are computed over two six-snapshot windows ($\sim 0.5$ Gyr each) for the two most massive samples used in this work: one between the time of infall and $\sim 0.5$ Gyr before infall, and another between $\sim 0.5$ and $\sim 1$ Gyr before infall. A positive value indicates that the corresponding property increases towards infall.}
\begin{tabular}{c c c c} 

  ${ 10^{11.5} < \mpeak}/\hMsun < 10^{12}$  & $\msub$  & $\Mstell$ & $\Mgas$ \\ [0.5ex] 
  $\rm t_0 \equiv t_{infall}$ & & & \\ [0.5ex] 
 \hline
 $\rm \langle \Delta m/ \Delta t\rangle\ {\rm [M_{\odot}/yr]}$ Centrals & 44.82 & 0.66 & 3.13  \\ 
 $\rm \langle \Delta m/ \Delta t\rangle\ {\rm [M_{\odot}/yr]}$ Satellites & 75.53 & 1.92 & -1.56  \\ 
 \\
   $\rm t_0 \equiv t_{infall} - 0.5\ Gyr$ & & & \\ [0.5ex] 
 \hline
 $\rm \langle \Delta m/ \Delta t\rangle\ {\rm [M_{\odot}/yr]}$ Centrals   & 46.25 & 0.68 & 3.38  \\ 
 $\rm \langle \Delta m/ \Delta t\rangle\ {\rm [M_{\odot}/yr]}$ Satellites & 79.80 & 1.95 & 3.17  \\ 
 \\
   ${ 10^{12} < \mpeak}/\hMsun < 10^{12.5}\ $  & $\msub$  & $\Mstell$ & $\Mgas$ \\ [0.5ex] 
  $\rm t_0 \equiv t_{infall}$ & & & \\ [0.5ex] 
 \hline
 $\rm \langle \Delta m/ \Delta t\rangle\ {\rm [M_{\odot}/yr]}$ Centrals & 135.16 & 2.44 & 5.38  \\ 
 $\rm \langle \Delta m/ \Delta t\rangle\ {\rm [M_{\odot}/yr]}$ Satellites & 266.83 & 6.23 & 2.01  \\ 
 \\
   $\rm t_0 \equiv t_{infall} - 0.5\ Gyr$ & & & \\ [0.5ex] 
 \hline
 $\rm \langle \Delta m/ \Delta t\rangle\ {\rm [M_{\odot}/yr]}$ Centrals   & 140.15 & 2.61 & 5.95  \\ 
 $\rm \langle \Delta m/ \Delta t\rangle\ {\rm [M_{\odot}/yr]}$ Satellites & 258.17 & 6.63 & 7.54 \\ 
\end{tabular}
\label{Table:mass}
\end{table}

In this appendix, we investigate the origin of the differences in stellar mass and gas mass between standard central galaxies and satellites shortly after infall. In Section~\ref{sec:mass_profile}, we find that central galaxies have less stellar mass and more gas than satellites shortly after infall, defined here as satellites with $\msub/\mpeak > 0.97$. These differences are particularly pronounced for the gas component, for which central galaxies contain substantial amounts of gas in the outer regions of their haloes, whereas satellites shortly after infall lack it.

We quantify these differences in Fig.~\ref{Fig:mpeak_rel_censat}, where we show the median stellar mass and gas fraction as a function of $\mpeak$. The black shaded region corresponds to the 16th and 84th percentiles of the distribution for central galaxies, and the coloured vertical shaded region corresponds to the $\mpeak$ selection used in this work. In terms of stellar mass, satellites shortly after infall have 40\%--60\% more stellar mass than the average central galaxy; at the same time, they can be up to 60\% less massive than the average satellite galaxy at $z=0$. For the gas fraction, the relative difference between central galaxies and satellites shortly after infall is about 20\%--50\%.

We identify two main causes for the differences in the properties of these two galaxy populations. The first is the effect of environment: satellite galaxies typically reside in denser environments than central galaxies. The second is that centrals shortly before infall already exhibit an evolutionary behaviour similar to that of satellite galaxies several hundred Myr before they are formally classified as satellites.

\subsection*{Large-scale environment}

Since more massive haloes exhibit a stronger cosmological bias, i.e. they reside in denser environments, and mergers occur more frequently in such environments, it is expected that satellite galaxies also inhabit denser environments than central galaxies of similar mass even before infall. The galaxy population differs across environments even at fixed halo mass, an effect known as occupancy variation. In \cite{Zehavi:2018} and \cite{Artale:2018}, the authors show that galaxies hosted by haloes of the same mass have higher stellar masses in denser environments, consistent with the results found in this paper. This effect is caused by the correlation between large-scale environment and secondary halo properties, such as halo concentration, which are themselves correlated with the galaxies they host; this is commonly referred to as halo assembly bias \citep{Gao:2005,Gao:2007, Contreras:2019}.

In the context of halo assembly bias, low-peak-height haloes, i.e. low-mass haloes at fixed redshift and cosmology, are expected to be more biased when they are more concentrated. This is shown in Fig.~\ref{Fig:hab}, which presents the halo bias measured in the \MTNGdmo\ simulation for all haloes, as well as for the upper and lower 20\% in concentration, in bins of halo mass. Following \cite{Contreras:2021_hab}, we compute the halo assembly bias as the ratio of the halo--matter cross-correlation function for haloes in a given mass bin to the theoretical matter correlation function. To minimise the impact of non-linearities and cosmic variance on the cross-correlation measurement, we compute the clustering on scales between 6 and 20 $\hMpc$ and average this ratio over those scales. The cross-correlation function was computed using the public code \texttt{CORRFUNC} \citep{corrfunc}, while the theoretical matter correlation function was obtained from the \texttt{HALOFIT} fitting function \citep{Takahashi:2012}, applied to the linear power spectrum from \texttt{CLASS} \citep{Lesgourgues:2011}, and then Fourier-transformed into configuration space.

As in the previous figures, we highlight in Fig.~\ref{Fig:hab} the halo-mass ranges that approximately correspond to the peak-subhalo-mass cuts used to classify galaxies in this work. While the halo mass in the gravity-only simulation does not map perfectly onto the peak subhalo mass of its hydrodynamic counterpart, we expect only minor differences in the halo selection for central galaxies. A halo with higher concentration is expected to host a central galaxy with higher stellar mass and lower star formation rate, which is related to its gas mass (see \citealt{Zehavi:2018} for a more detailed discussion). We also note that, in agreement with the differences between central galaxies and satellites shortly after infall found in this work, the differences between the high- and low-concentration samples decrease with increasing halo mass.

To quantify the impact of the large-scale environment on the properties of central galaxies, we show in Fig.~\ref{Fig:mpeak_rel_env} the median stellar mass and gas mass normalised by peak subhalo mass as a function of peak subhalo mass for central galaxies living in the 20\% densest and least dense environments. To compute the environment, we use the individual bias-per-object methodology \citep{Paranjape:2018}, which effectively measures the bias of an individual object. In agreement with expectations, we find that centrals in denser environments have more stellar mass and less gas than centrals in less dense environments, although the environmental effect is weaker than the difference shown in Fig.~\ref{Fig:mpeak_rel_censat}. The same conclusion can be reached from Fig.~\ref{Fig:profile_centrals}, which shows the mass profiles of centrals living in the densest and least dense environments. Again, centrals in denser environments have more stellar mass and less gas than those in less dense environments. The differences are concentrated mainly around the peaks of each profile and in the outer regions, similarly to the differences found between central galaxies and satellites shortly after infall, although with a smaller amplitude (Fig.~\ref{Fig:profile}). It is possible that the environment is not itself the direct cause, but rather traces another effect that drives part of this difference, or that it is only partially responsible for the trends found in this work. That said, among the two baryonic quantities, stellar mass appears to be more strongly affected by environment than gas mass, whereas gas shows larger differences between central galaxies and satellites shortly after infall. This may indicate that environmental effects are not the only drivers of these differences.

\subsection*{Pre-evolution of central galaxies}

Another effect responsible for the differences between central galaxies and satellites shortly after infall is the evolution of central galaxies before they become satellites. If centrals begin to be affected by a nearby halo before the time they are formally identified as satellites, it is possible that both their mass components and their internal mass distribution have already changed significantly by the time of infall. To test this hypothesis, we measure the times at which satellite galaxies reach their peak subhalo mass (sometimes referred to as the time of $\mpeak$), stellar mass, and gas mass, and compare them with the time of infall (Table~\ref{Table:times}). We find that, for the two most massive samples considered in this work, both the peak subhalo mass and the peak gas mass are reached before the galaxy becomes a satellite. This implies that mass loss begins before the galaxy is formally classified as a satellite. It is important to note that the time at which a galaxy reaches its peak gas mass precedes the time at which it reaches its peak subhalo mass. This picture suggests that, as a galaxy approaches a more massive halo and begins to interact with the outer regions of the host halo, it can still increase its total mass through accretion while its gas is already being stripped.

To build a more general picture of the evolution of centrals before they become satellites, we measure the change in mass as a function of time for both central and satellite galaxies (Table~\ref{Table:mass}). The mass changes are computed over two six-snapshot windows ($\sim 0.5$ Gyr each): one between the time of infall and $\sim 0.5$ Gyr before infall, and another between $\sim 0.5$ and $\sim 1$ Gyr before infall. We find that satellite galaxies show a larger increase in total mass than central galaxies, by about 60\%--100\%, in both time windows and across both galaxy samples considered. The stellar mass also shows a larger increase, even more pronounced than for the total mass, by about 250\%--280\%. This suggests that interaction with the nearby halo may trigger the conversion of cold gas into stars, which could also help to explain the differences found between central galaxies and satellites shortly after infall. Finally, for the gas component, we find that its evolution changes significantly as infall approaches. During the period between 1 Gyr and 0.5 Gyr before infall, the increase in gas mass is relatively similar for centrals and satellites, although this still corresponds to only a modest increase for satellites, since during the same period their subhalo masses grow faster than those of central galaxies. In the last 0.5 Gyr before infall, the gas mass barely increases for the more massive satellite sample, while the less massive sample already begins to lose gas. These trends are the same as those found for satellite galaxies elsewhere in this work, strongly suggesting that centrals begin to behave like satellites well before they are formally classified as such. We would like to highlight that these mass variations exhibit large scatter and that these numbers should be used only to understand satellite evolution, not for modelling unless a more dedicated study is performed.

We find that both the environmental differences and the distinct pre-infall evolution of centrals shortly before becoming satellites are linked to the differences between central galaxies and satellites shortly after infall. The results presented here, however, cannot disentangle the correlation between these two effects, nor determine whether other processes also contribute to these differences. A more extensive study of this issue would be of clear interest, but lies beyond the scope of this paper. In future work, we plan to investigate the main cause of these differences in more detail. The conclusions of such a study would not only improve our understanding of galaxy formation physics, but also help refine the construction of mock catalogues and potentially clarify some aspects of halo assembly bias.

\section{Additional parameterisations}
\label{sec:add_param}

\begin{figure*}
\centering
\includegraphics[width=0.245\textwidth]{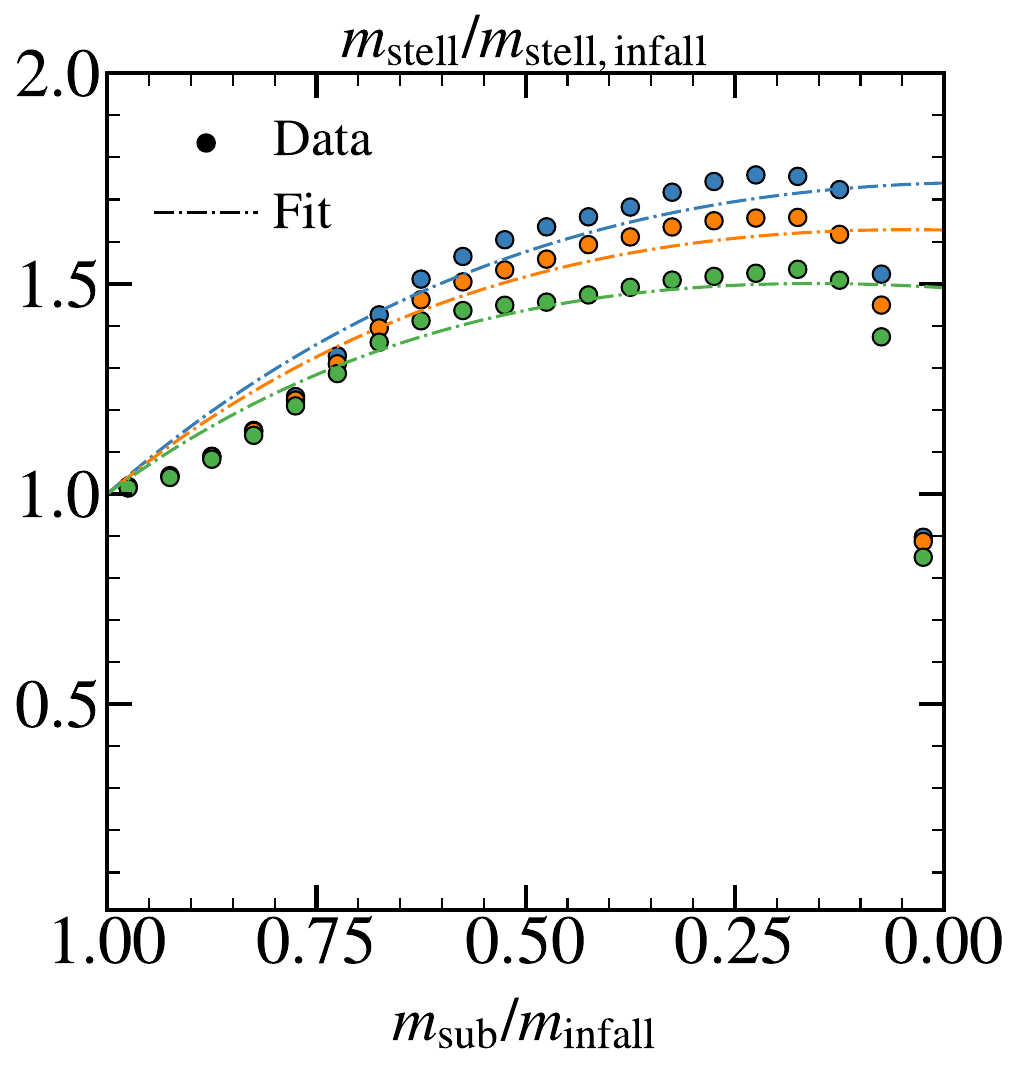}
\includegraphics[width=0.245\textwidth]{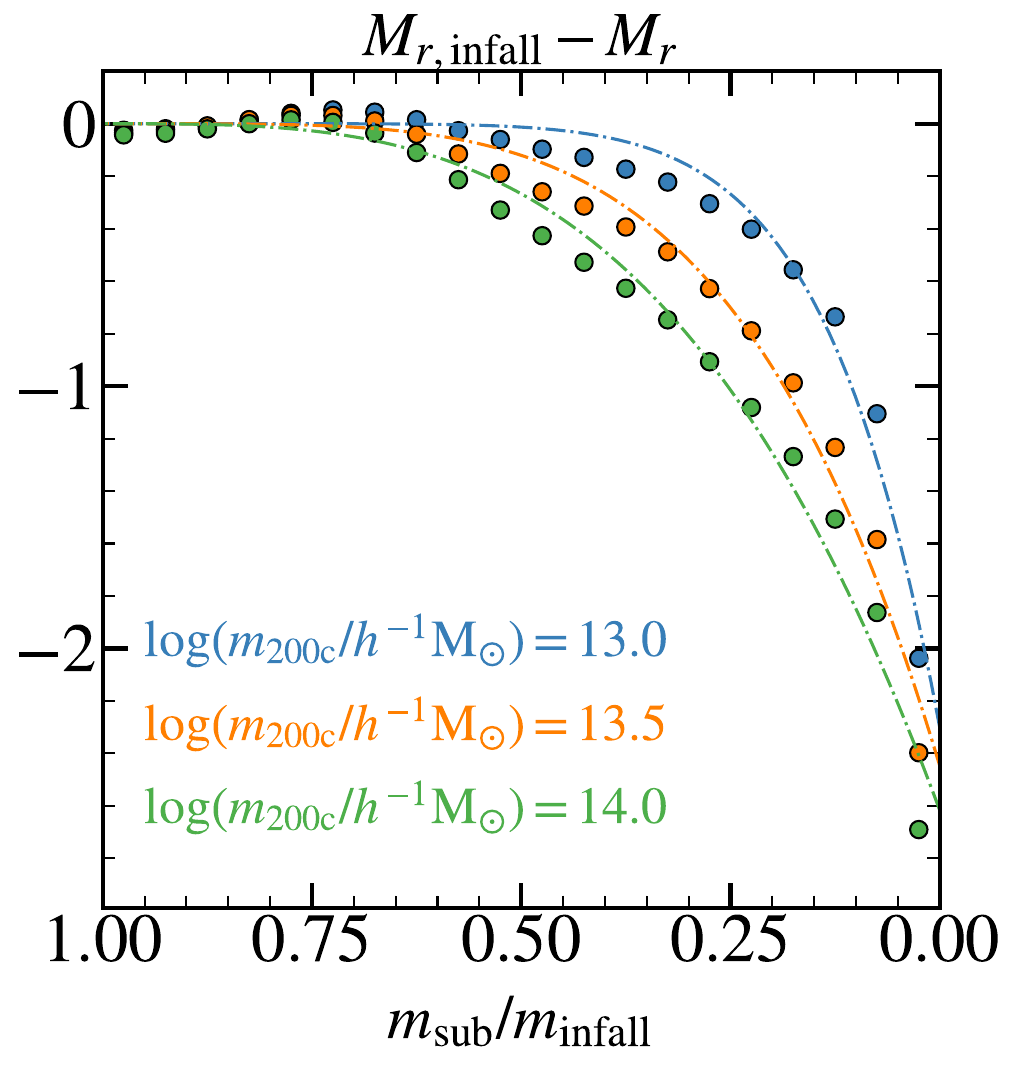}
\includegraphics[width=0.245\textwidth]{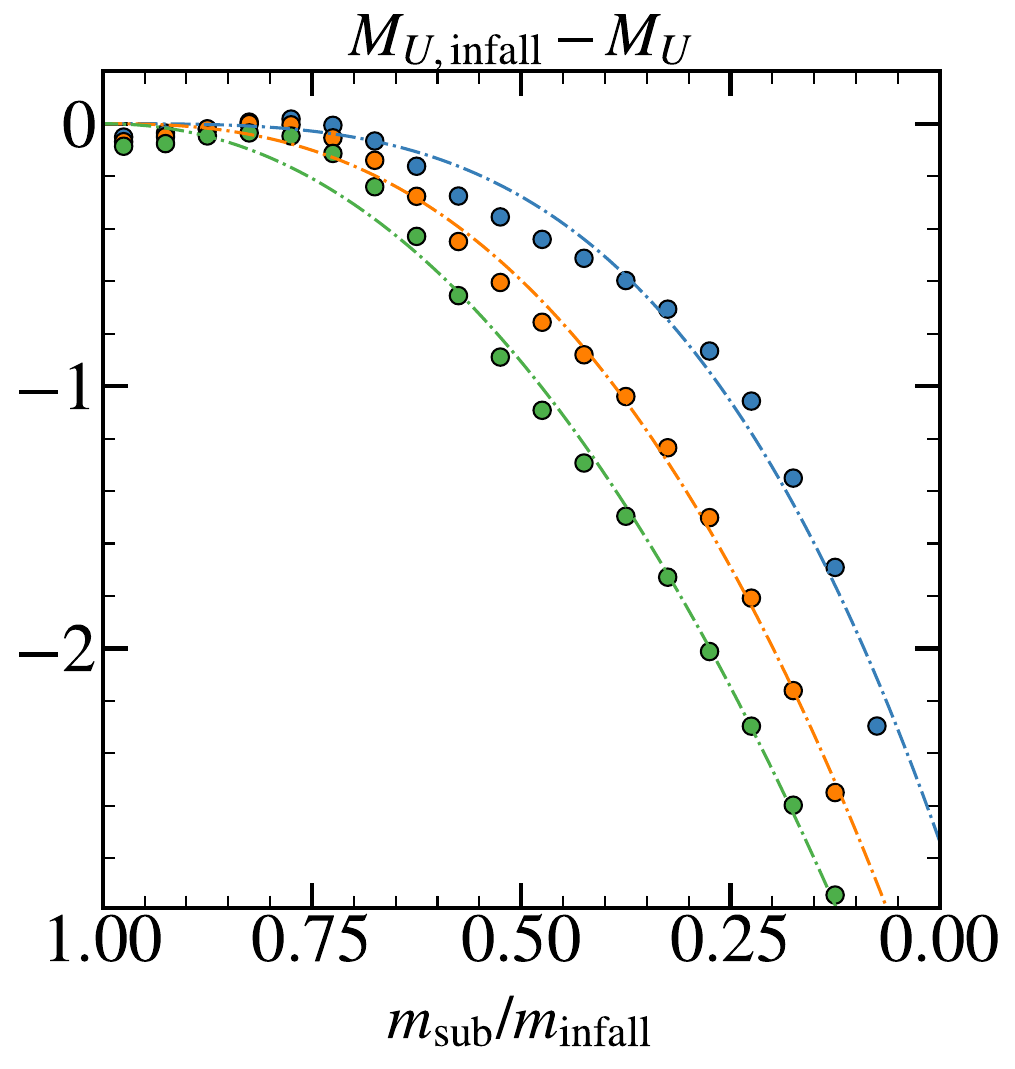}
\caption{Similar to Fig.~\ref{fig:fits}, but showing the evolution of the stellar mass, and $r$- and $U$-band magnitudes as a function of $\msub/\minfall$ for galaxies with $11.5 < \log(\mpeak/\hMsun) < 12.0$. The symbols represent the data from the \MTNG\ simulation, as shown in Fig.~\ref{Fig:gen_ev}, while the dash-dotted lines represent the analytic models described in Section~\ref{sec:EvBar_fit}. The different colours represent different host halo masses, as labelled.}
\label{Fig:Fit_msub}
\end{figure*}

\begin{figure*}
\includegraphics[width=0.245\textwidth]{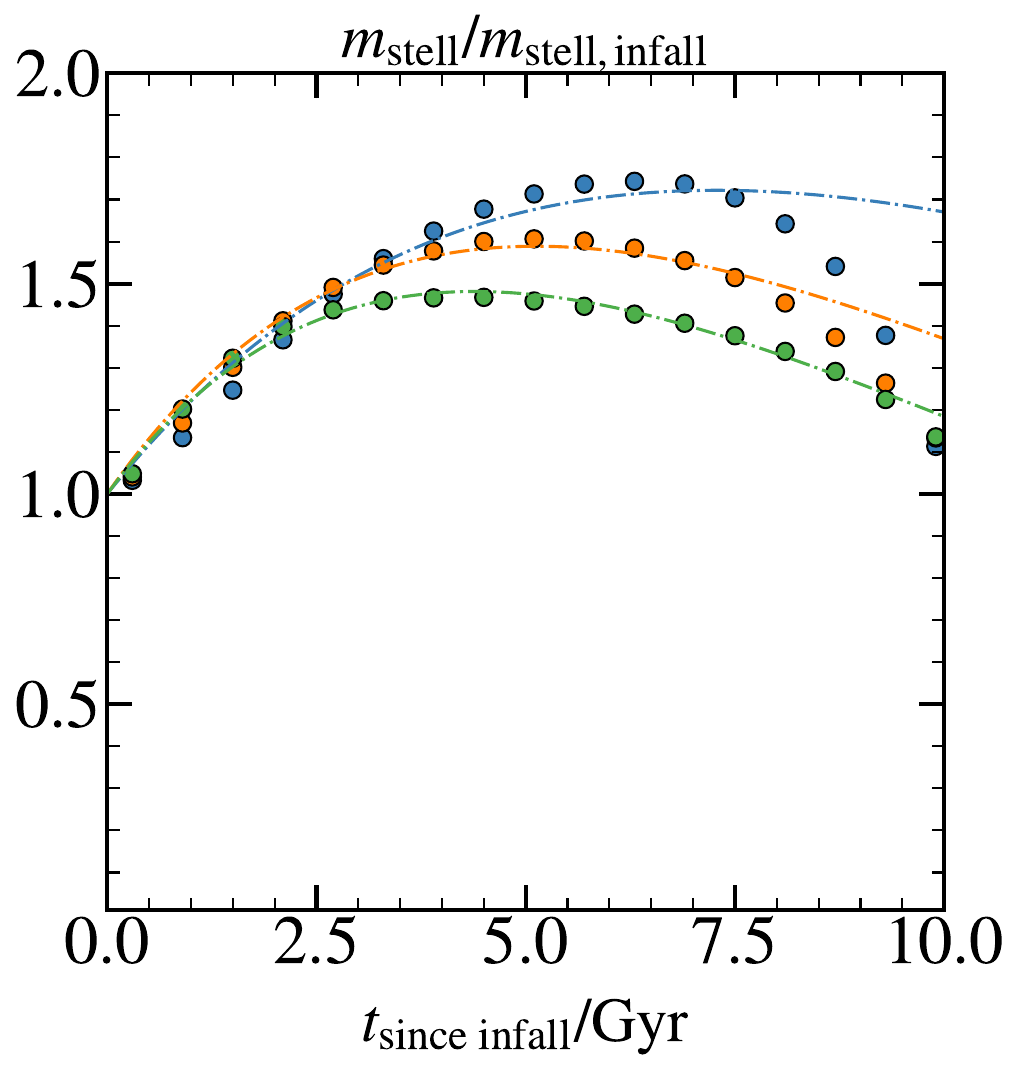}
\includegraphics[width=0.245\textwidth]{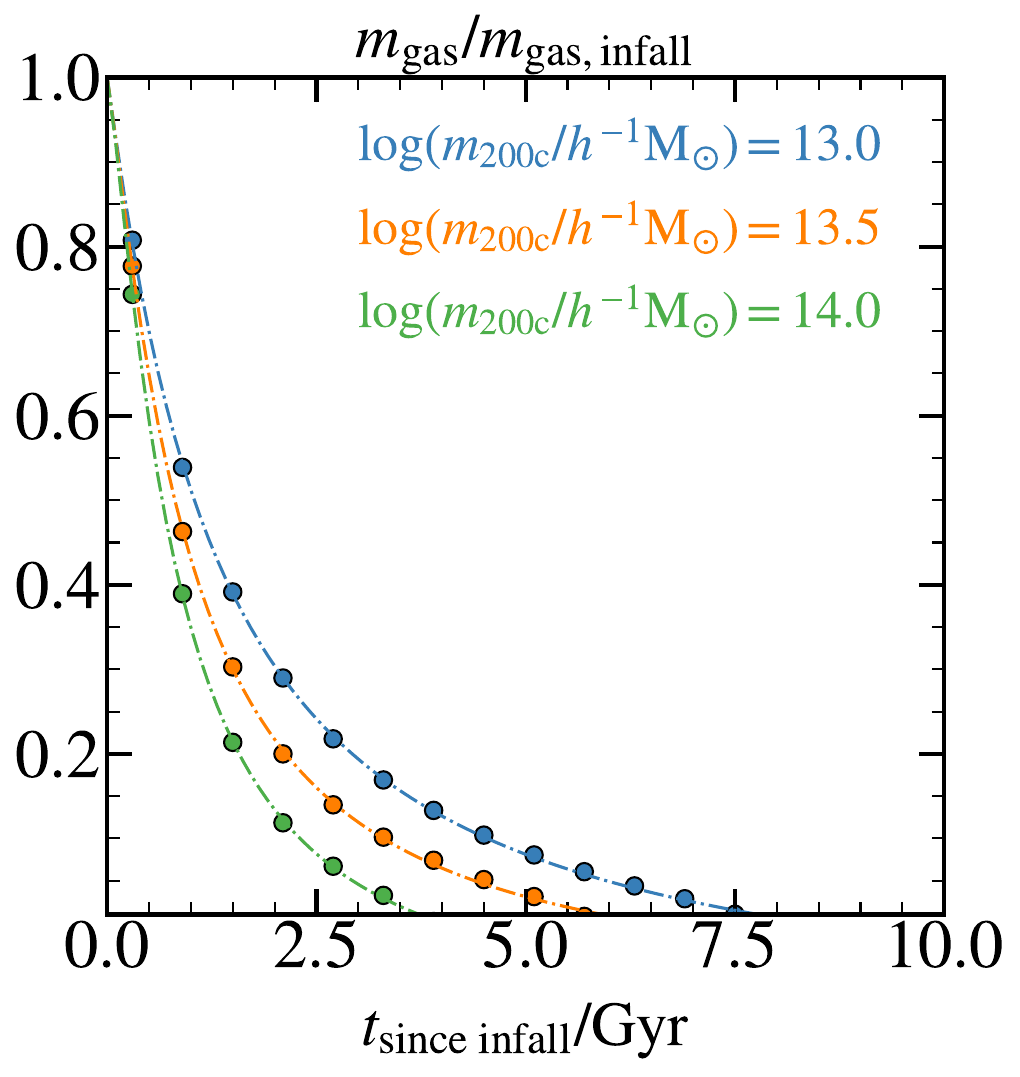}
\includegraphics[width=0.245\textwidth]{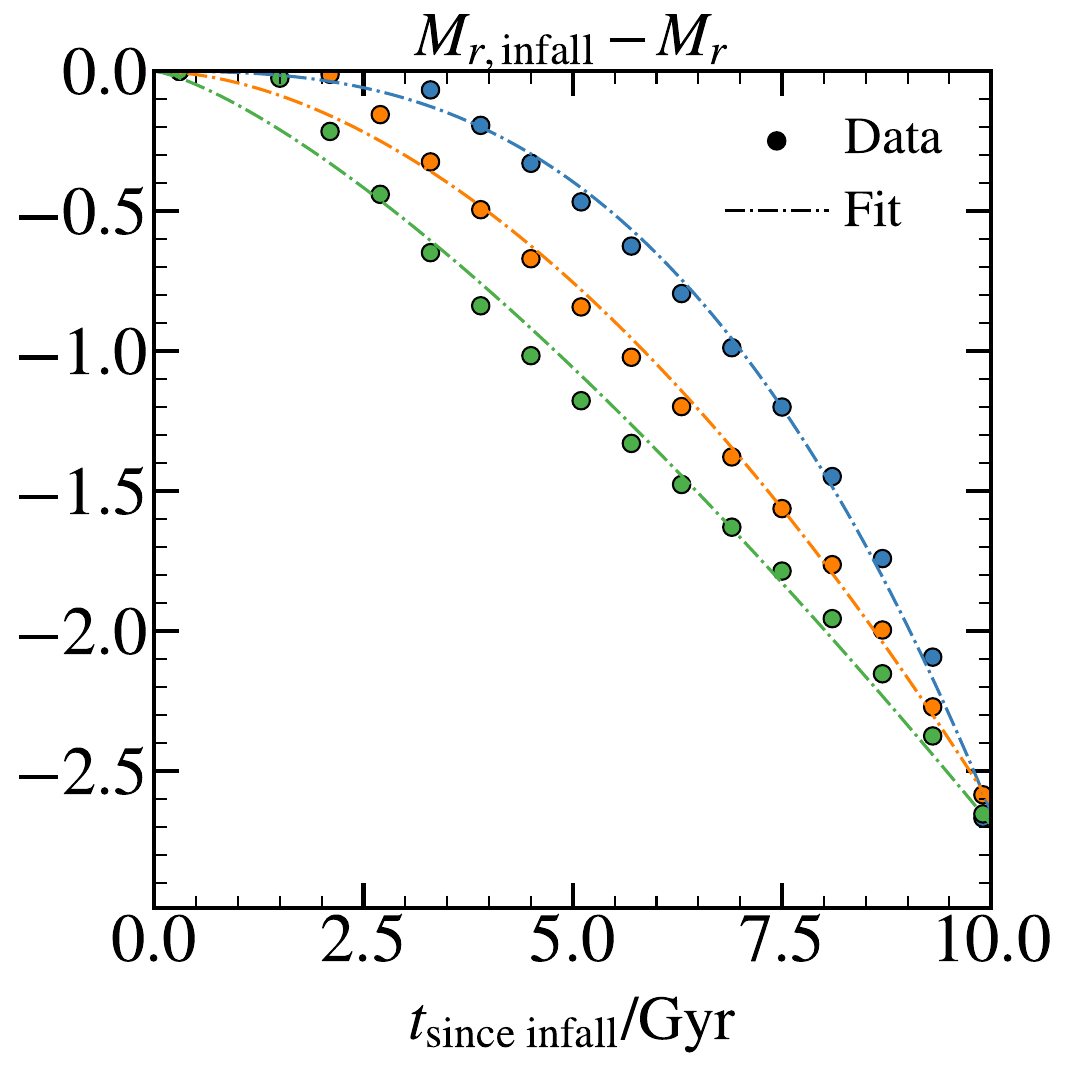}
\includegraphics[width=0.245\textwidth]{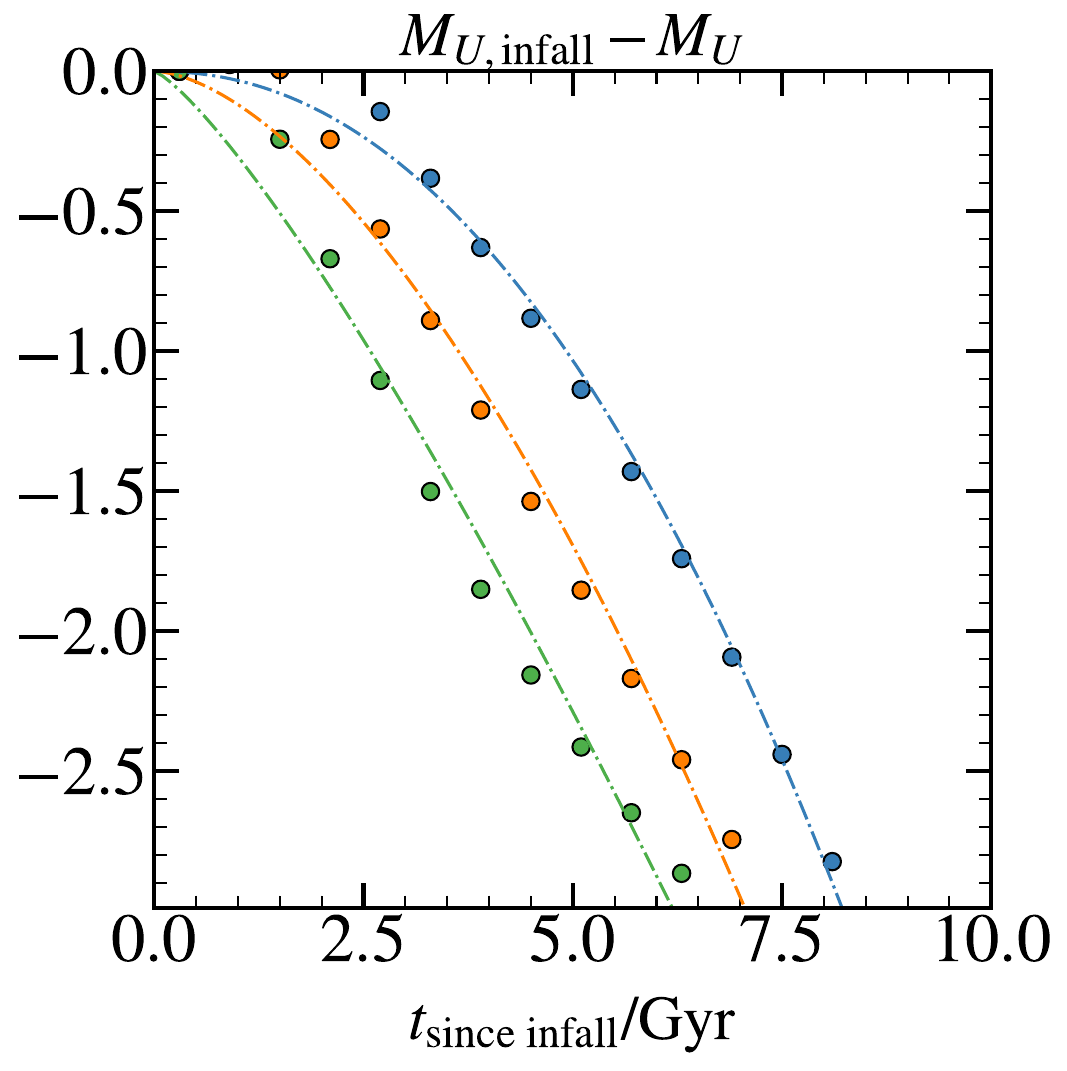}
\caption{Similar to Fig.~\ref{fig:fits}, but showing the evolution of the stellar mass, gas mass, and $r$- and $U$-band magnitudes as a function of time since infall for galaxies with $11.5 < \log(\mpeak/\hMsun) < 12.0$. The symbols represent the data from the \MTNG\ simulation, as shown in Fig.~\ref{Fig:gen_ev}, while the dash-dotted lines represent the analytic models described in Section~\ref{sec:EvBar_fit}. The different colours represent different host halo masses, as labelled.}
\label{Fig:Fit_tinfall}
\end{figure*}

\begin{figure*}
\includegraphics[width=0.245\textwidth]{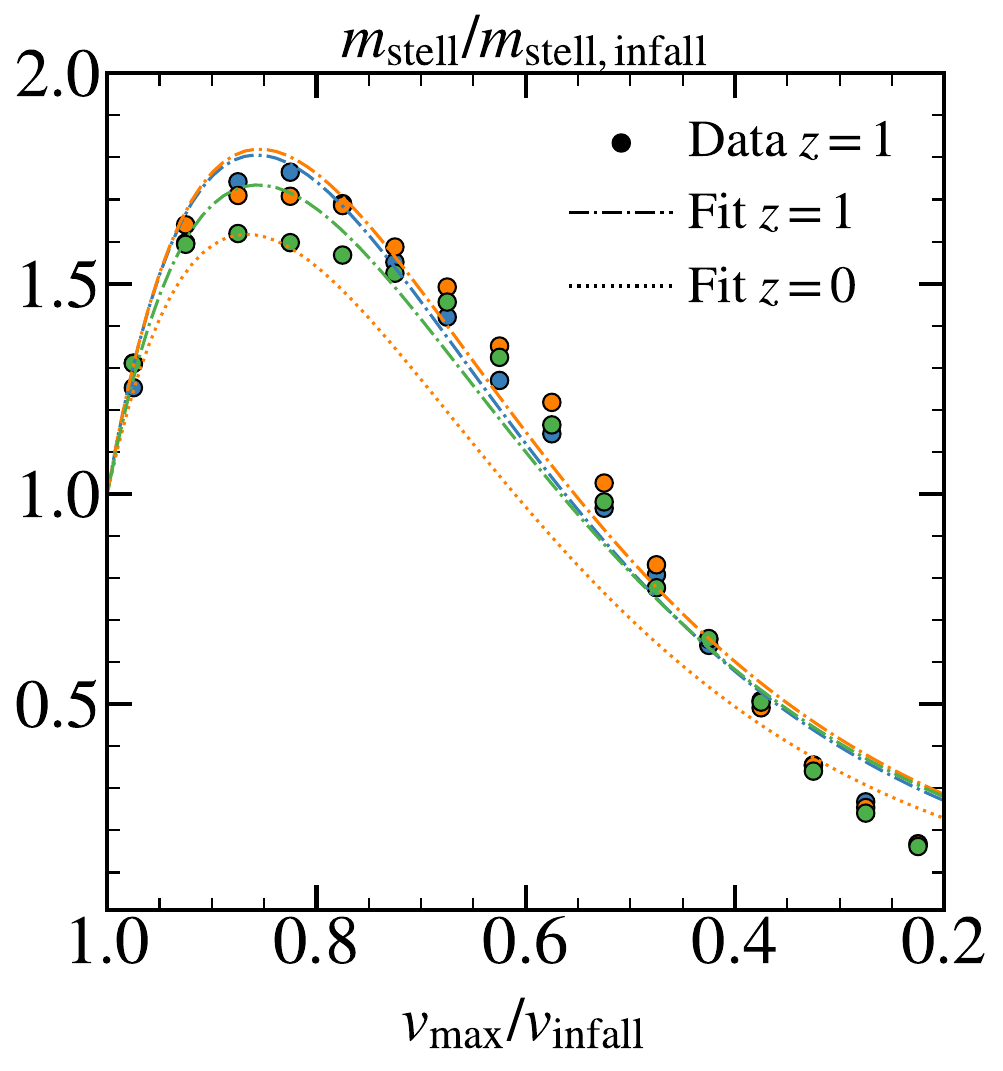}
\includegraphics[width=0.245\textwidth]{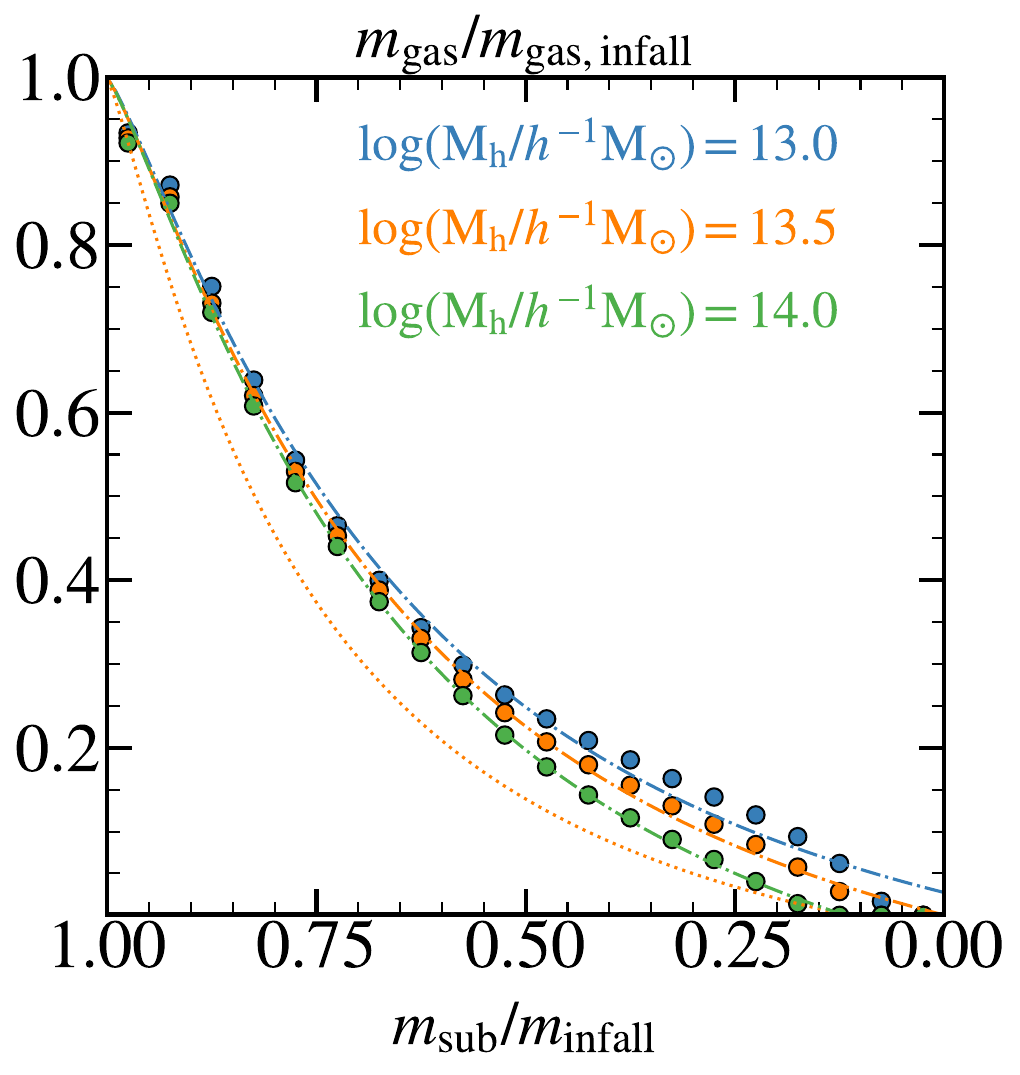}
\includegraphics[width=0.245\textwidth]{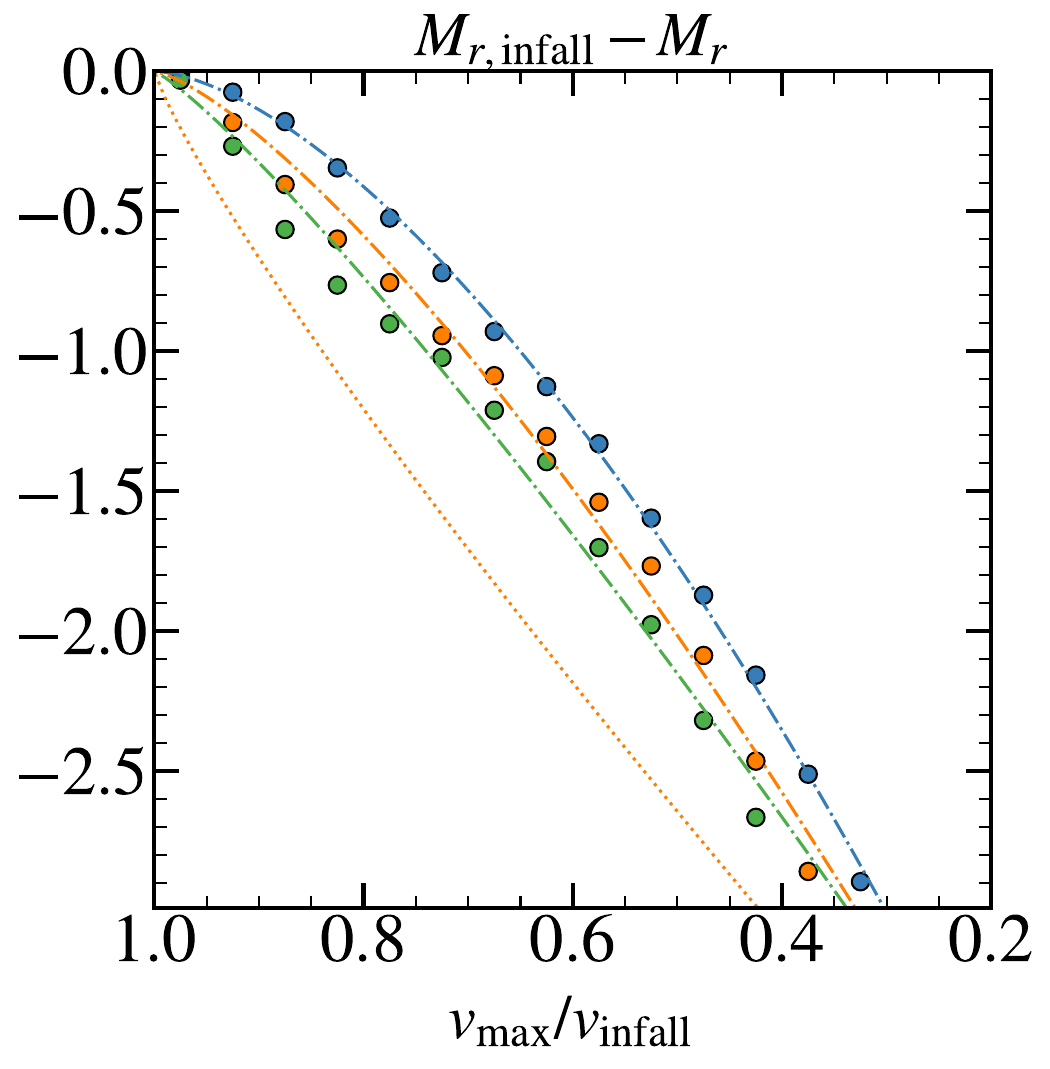}
\includegraphics[width=0.245\textwidth]{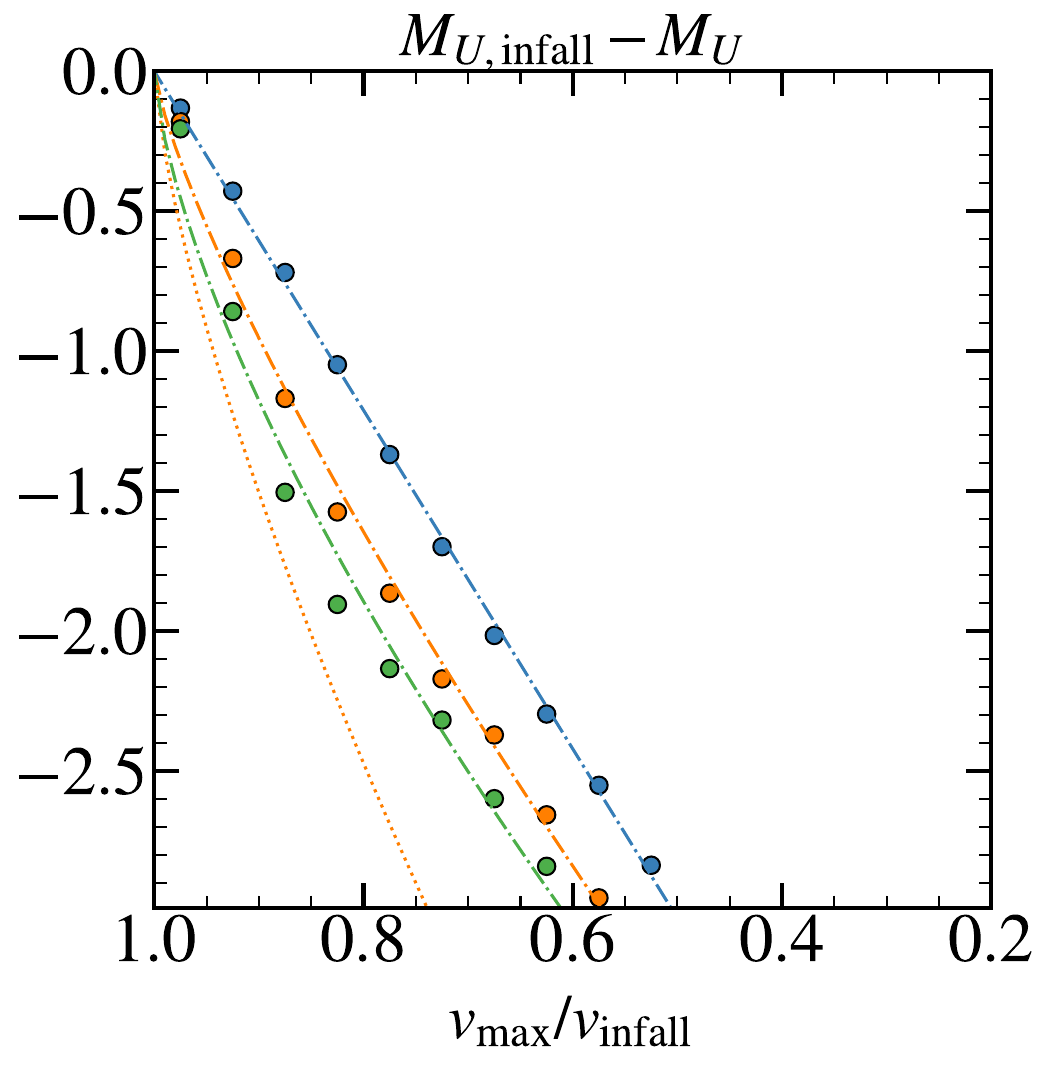}
\caption{Similar to Fig.~\ref{fig:fits}, but showing the evolution of the stellar mass as a function of $\vmax/\vinfall$, the gas mass as a function of $\msub/\minfall$, and the $r$- and $U$-band magnitudes as a function of $\vmax/\vinfall$ for galaxies with $11.5 < \log(\mpeak/\hMsun) < 12.0$ at $z=1$. The symbols represent the data from the \MTNG\ simulation, as shown in Fig.~\ref{Fig:gen_ev}, while the dash-dotted lines represent the analytic models described in Section~\ref{sec:EvBar_fit}. The different colours represent different host halo masses, as labelled. As a reference, the orange dotted line shows the evolution of the intermediate-halo mass sample at $z=0$.}
\label{Fig:Fit_z1}
\end{figure*}

\begin{figure*}
\includegraphics[width=0.245\textwidth]{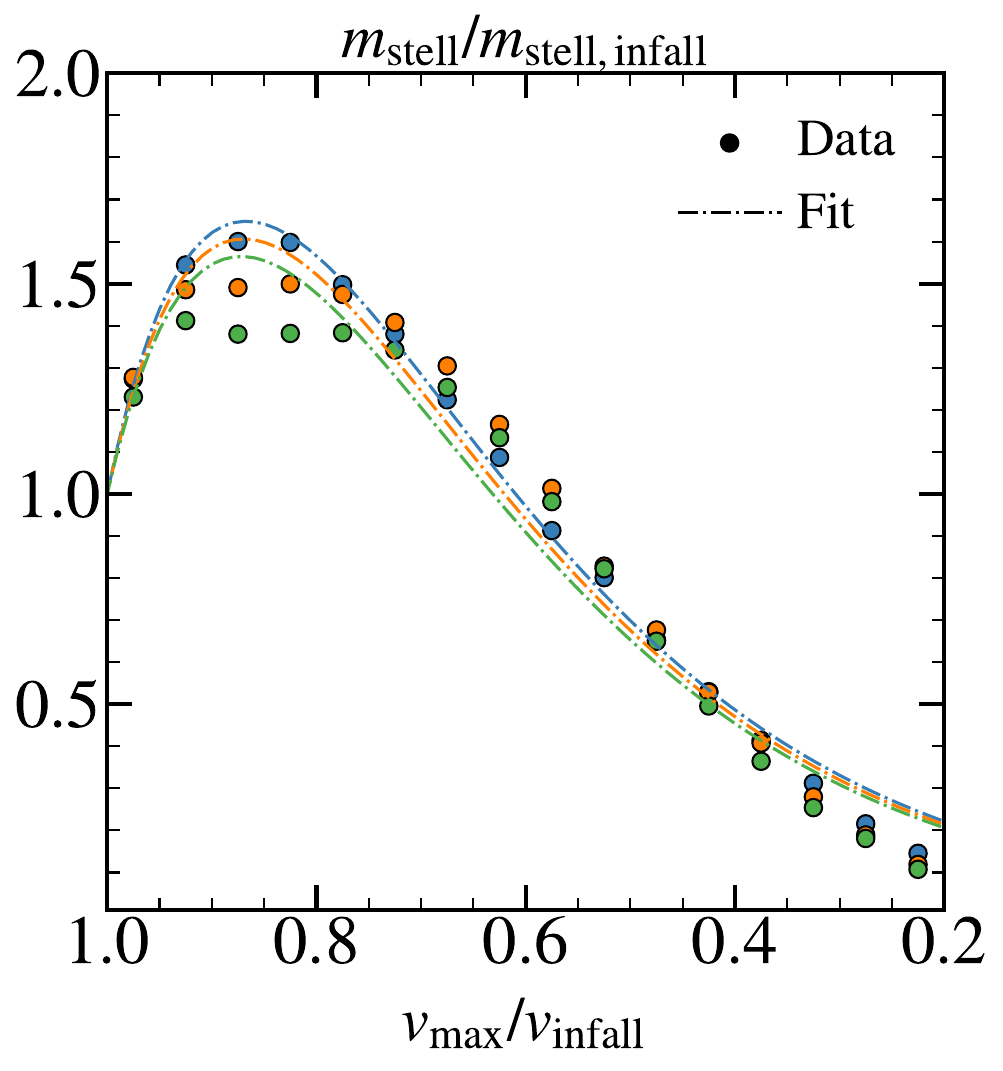}
\includegraphics[width=0.245\textwidth]{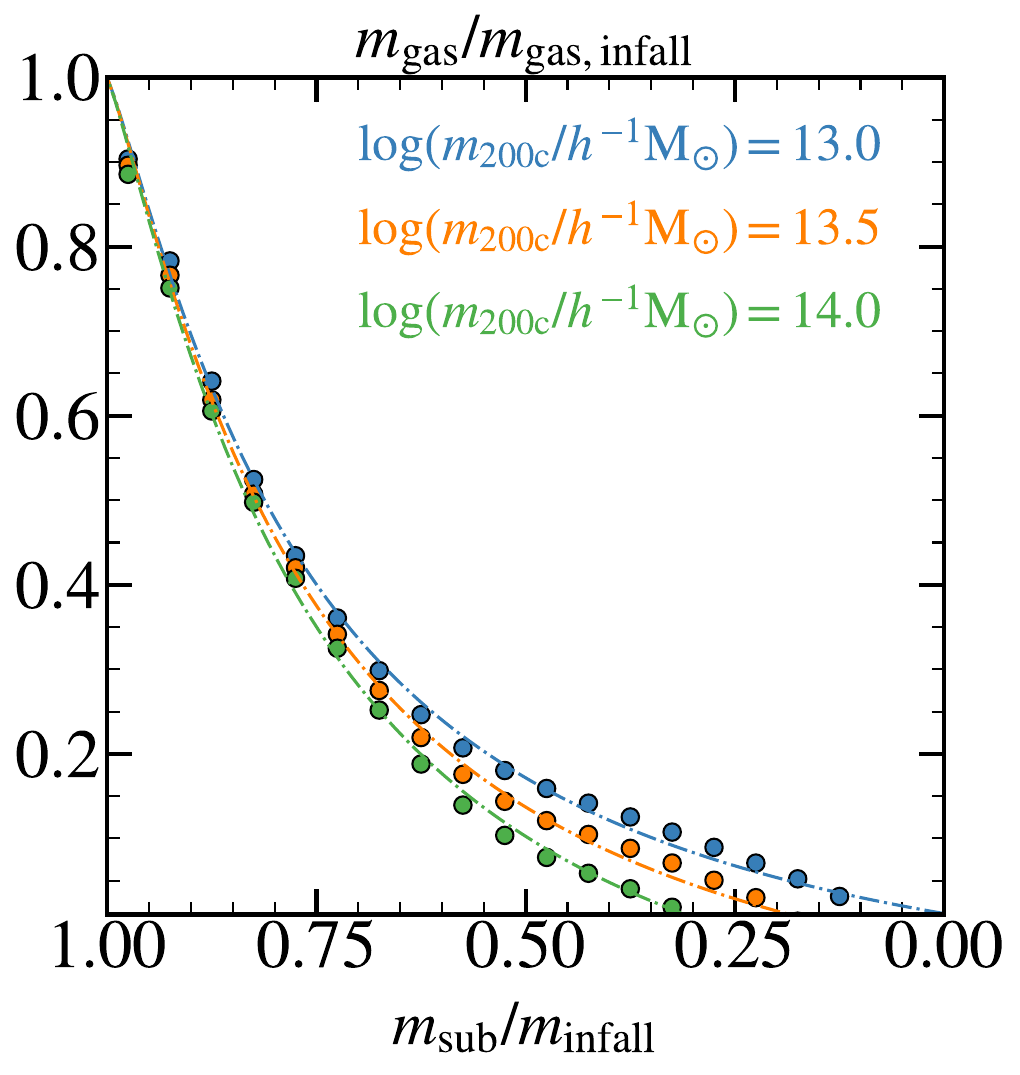}
\includegraphics[width=0.245\textwidth]{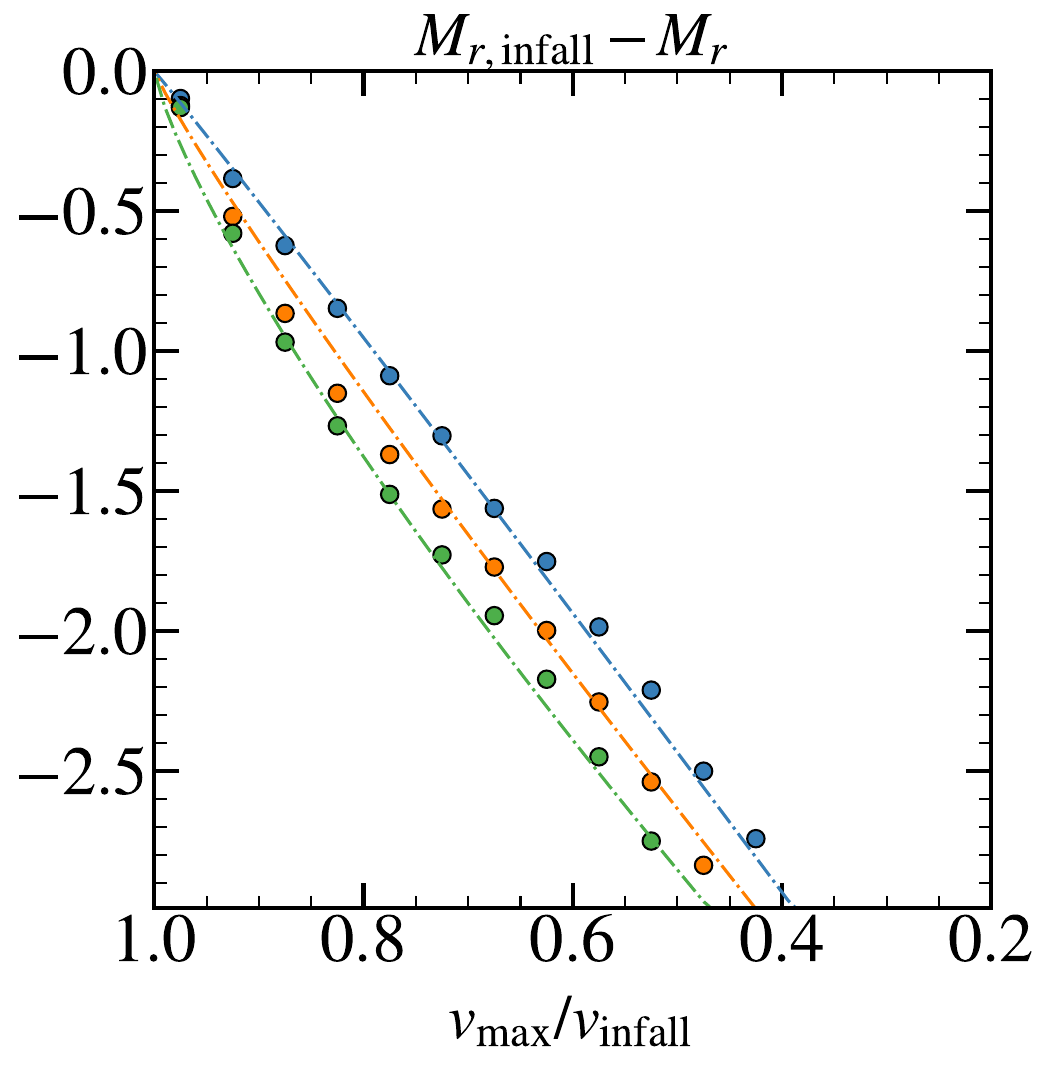}
\includegraphics[width=0.245\textwidth]{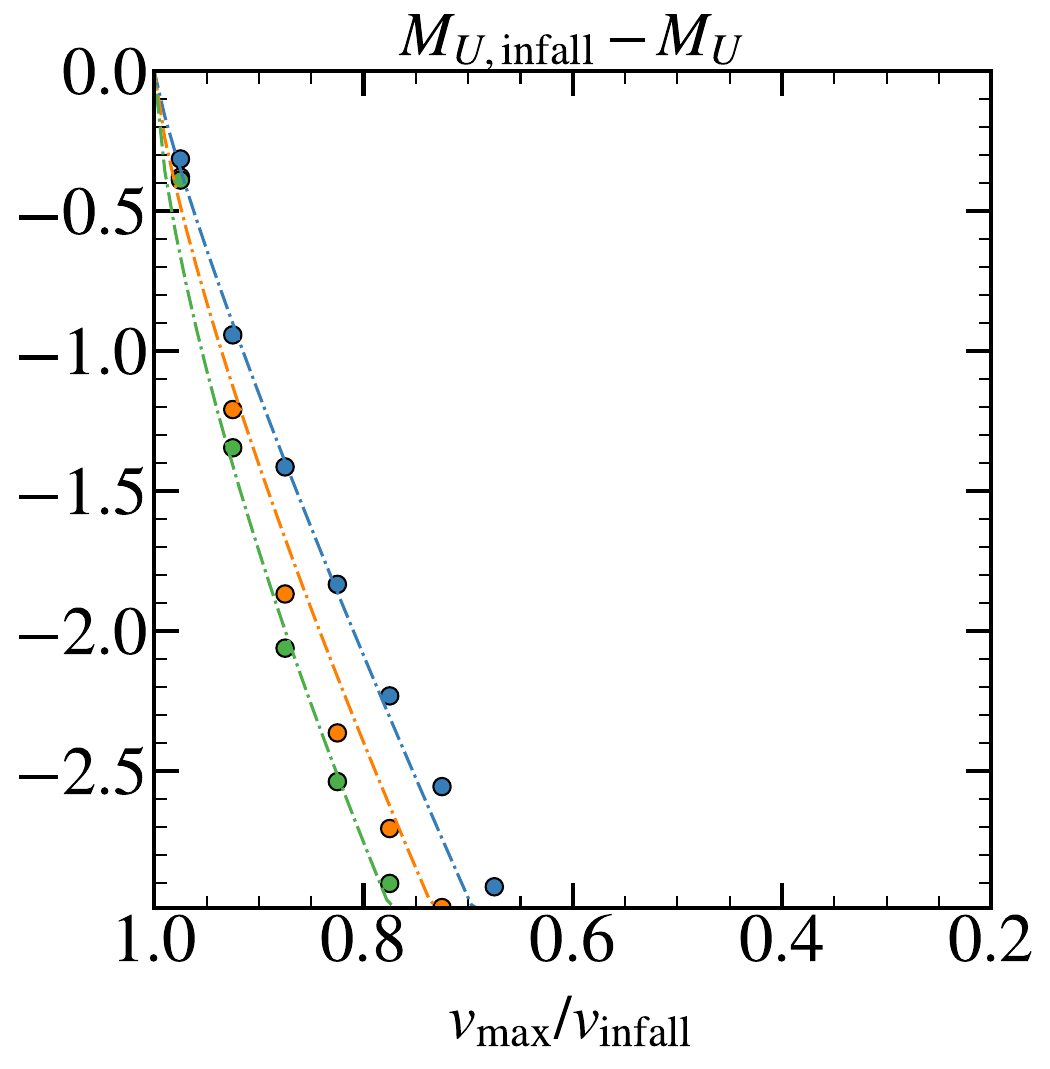}
\caption{The evolution of the stellar mass as a function of $\vmax/\vinfall$, the gas mass as a function of $\msub/\minfall$, and the $r$- and $U$-band magnitudes as a function of $\vmax/\vinfall$ for galaxies with $11.5 < \log(\mpeak/\hMsun) < 12.0$. The different colours represent different host halo masses, as labelled. The symbols represent the data from the \MTNG\ simulation, while the dash-dotted lines represent an analytic model fitted simultaneously to all halo masses.}
\label{Fig:Fit_single}
\end{figure*}

\begin{figure}
\centering
\includegraphics[width=0.45\textwidth]{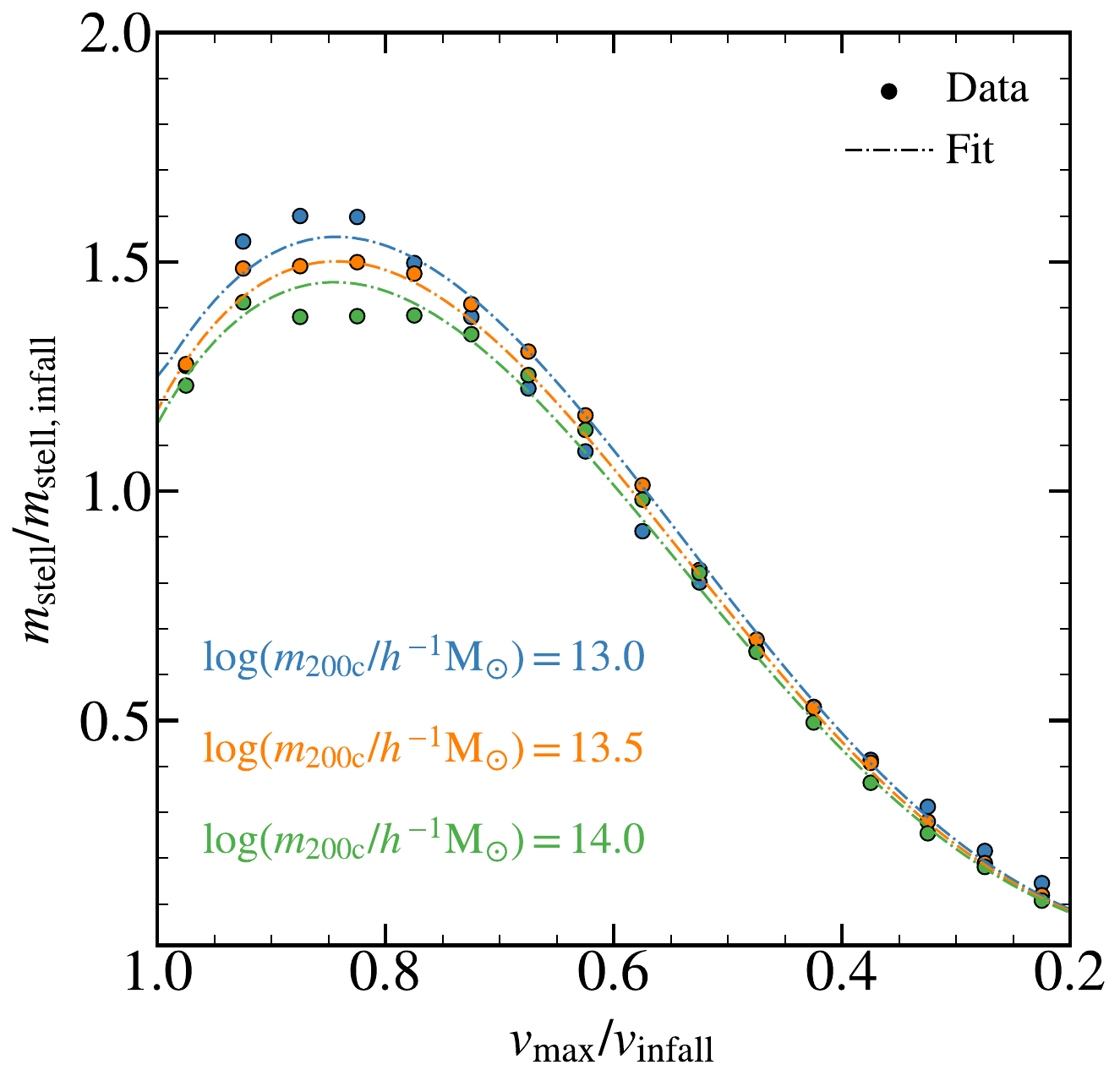}
\caption{The evolution of stellar mass as a function of $\vmax/\vinfall$. The dash-dotted line represents an analytic model, obtained through symbolic regression, fitted simultaneously to all halo masses.}
\label{Fig:Fit_LR}
\end{figure}

In this section, we show how the parameterisations presented in Section~\ref{sec:EvBar_fit} perform when other subhalo quantities are used. In Fig.~\ref{Fig:Fit_msub}, we show the evolution of the stellar mass and the $r$- and $U$-band magnitudes as a function of $\msub/\minfall$ for galaxies with $11.5 < \log(\mpeak/\hMsun) < 12.0$. We do not include the evolution of the gas mass, since that property is already shown in Fig.~\ref{fig:fits}. The remaining subhalo mass fraction is more commonly used to describe satellite evolution, as it tends to be more readily available than $\vmax/\vinfall$. We find that, by using the same parameterisation as for $\vmax/\vinfall$, we can recover the evolution of most properties almost perfectly. The only property that presents some difficulty in the modelling is the stellar mass. In this case, we are not able to reproduce the abrupt decrease in stellar mass that occurs when $\msub/\minfall \sim 0.1$. Although such a feature could be expected \citep{Errani:2022}, we warn the reader about possible contamination in the calculation of the remaining subhalo mass due to the presence of stellar particles in the core of the galaxy, which are absent in gravity-only simulations.

We then test the parameterisations when modelling the evolution of galaxy properties as a function of time since infall. This quantity does not share the limitation of being computed from the mass distribution within the subhalo, as is the case for $\msub/\minfall$ and $\vmax/\vinfall$, which may differ between hydrodynamic and gravity-only simulations because, in the former, baryonic particles and cells, especially stellar particles, contribute to the inner mass distribution of haloes, whereas this component is absent in gravity-only simulations. We show the evolution of the galaxy properties as a function of time since infall, together with their respective parametric forms, in Fig.~\ref{Fig:Fit_tinfall}. We again use the parameterisations described in Section~\ref{sec:EvBar_fit}, although we redefine the fitting variable as $x \equiv 1 - t_{\rm since\ infall}/(10\,{\rm Gyr})$. Similar to Fig.~\ref{Fig:Fit_tinfall}, we can successfully model the evolution of the $r$- and $U$-band magnitudes, as well as the gas mass. We only struggle to reproduce the final stage of the stellar-mass evolution, for which more complex modelling may be required.

In this paper, we focus mostly on the evolution of satellite galaxies at $z=0$. Nonetheless, Fig.~\ref{Fig:gen_ev_z1} shows that the evolution of galaxy properties at $z=1$ is similar to that at $z=0$. In Fig.~\ref{Fig:Fit_z1}, we show that the parameterisations used throughout this paper remain valid at $z=1$ by fitting the evolution of galaxy properties at this redshift. As a reference, we also show the parameterisation of the evolution of galaxy properties for the intermediate-host-halo-mass sample at $z=0$. Although the evolution at the two redshifts is not identical, both follow the same trends and remain sufficiently similar to suggest that the parameterisations presented in this work could be used for satellites identified at different redshifts.

The final set of parameterisations we consider uses the same combination of galaxy and subhalo properties as Fig.~\ref{fig:fits}, but adopts a common parametric form for all host halo masses. For this fit, we add one free parameter to each parameterisation, modifying the values of $\gamma$, $A$, and $B$ in equations (1), (2), and (3), respectively. The new model assumes a linear dependence of these parameters on the logarithm of the halo mass, such that $P = P_0 + dP\,[13 - \log(\mvir/\hMsun)]$, where $P_0$ is the original free parameter and $dP$ is a new free parameter that captures the dependence on halo mass. As shown in Fig.~\ref{Fig:Fit_single}, all galaxy properties can be well reproduced at all halo masses with this new parameterisation. Although promising, a wider halo-mass range must be explored to demonstrate properly that this parameterisation can be applied across the full halo population of the simulation.

Regarding the fit to the stellar-mass evolution, we also test symbolic regression in an attempt to provide a more accurate parameterisation of the data, even if it lacks the physical motivation of the form used throughout this work. We find that a parameterisation of the form:

\begin{equation}
f(x) = \frac{A\,x}{\left[-\exp(Bx) + \ln\!\left(\log(M)\right)\right]^{\,C-Dx}}
\end{equation}

\noindent where $x \equiv \vmax/\vinfall$ and $A$, $B$, $C$, and $D$ are free parameters. This parameterisation can reproduce the data almost perfectly, as shown in Fig.~\ref{Fig:Fit_LR}, but its highly flexible form strongly suggests that it would struggle to generalise to other implementations. In addition, because we do not force it to take a value of unity when $\vmax/\vinfall = 1$, it may present difficulties when implemented in gravity-only simulations. We nevertheless include this parameterisation to illustrate the potential of symbolic regression to provide highly accurate fits in cases where reproducing a specific model is the main goal.

\section{Assessing the impact of the SUBFIND-HBT gas-assignment bias on satellite gas masses}

\begin{figure}
\centering
\includegraphics[width=0.45\textwidth]{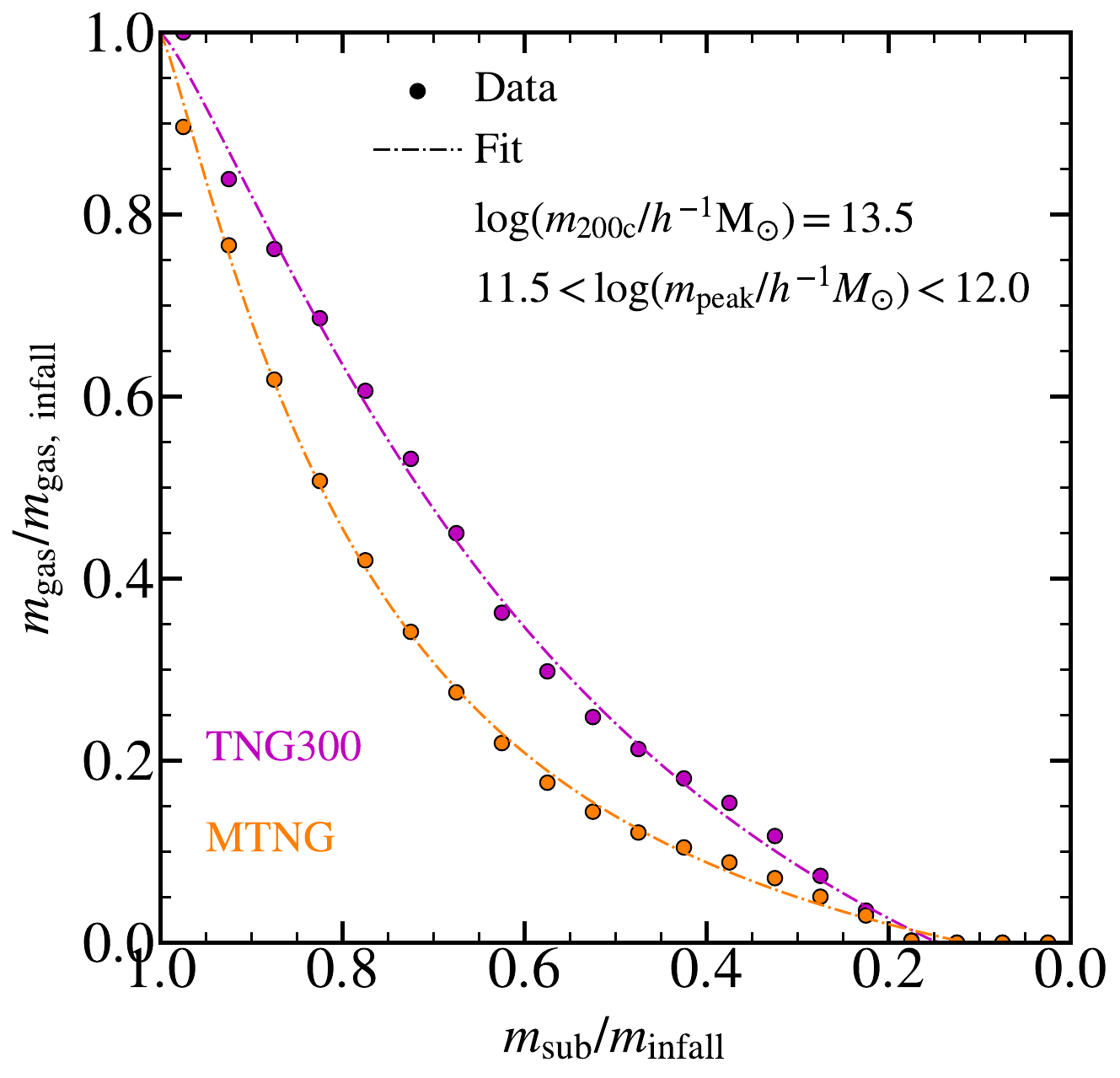}
\caption{Similar to the top right panel of Fig.~\ref{fig:fits}, but showing only the galaxy sample with $\log(\mvir/\hMsun) = 13.5$ and including in addition the predictions from the TNG300 simulation. The symbols represent the data from the simulations, while the dash-dotted lines represent the analytic models described in Section~\ref{sec:EvBar_fit}.}
\label{Fig:TNG300_z0}
\end{figure}

\label{sec:SUBFIND-HBT}
{In this appendix, we provide an approximate assessment of the impact of the gas-cell assignment procedure used by the version of SUBFIND-HBT adopted in MTNG. As discussed in Section~\ref{sec:sims}, this implementation can systematically underestimate the diffuse gas mass associated with satellite subhaloes. To estimate the magnitude of this effect, we compare the gas-content trends measured in MTNG with those measured in TNG300, which is not affected by this issue in the same way.

The TNG300 simulation has a periodic box of 205 $\hMpc$ ($\sim 300$ Mpc) on a side and contains $2500^3$ dark matter particles and an equal number of gas cells, implying a baryonic mass resolution of $7.44\times10^6\,\hMsun$ and a dark matter particle mass of $3.98\times 10^7\,\hMsun$. Its outputs are publicly available at \url{www.tng-project.org/data} \citep{Nelson:2019}. The differences in volume, resolution, and hydrodynamic implementation mean that the comparison between the two simulations can only be used as an approximate diagnostic and should not be interpreted as a full validation. Nevertheless, it provides a useful way to obtain a rough estimate of the amplitude of the bias introduced by the MTNG implementation.

In Fig.~\ref{Fig:TNG300_z0}, we compare the evolution of the gas mass as a function of $\msub/\minfall$ in the MTNG and TNG300 simulations, similarly to Figs.~\ref{Fig:gen_ev} and~\ref{fig:fits}. We find that the gas masses associated with satellites in MTNG are systematically lower than those measured in TNG300, consistent with the expected bias introduced by the MTNG implementation of SUBFIND-HBT. However, the differences are not catastrophic and are of similar magnitude to the scatter shown in Fig.~\ref{Fig:gen_ev}. In addition, the overall shape of the relation is broadly similar in the two simulations, and Eq.~(1) is also capable of characterising the evolution of the gas mass in TNG300 (magenta dash-dotted line). We also examine the gas-mass profiles of satellite galaxies in TNG300 and find qualitatively similar behaviour to that seen in MTNG, in the sense that gas loss does not proceed strictly from the outside in (not shown here).

These comparisons suggest that the main results of this work concerning the evolution of satellite gas content remain qualitatively robust, in the sense that the overall trend appears to be preserved. At the same time, it indicates that the absolute gas-mass scale inferred from MTNG should be interpreted with caution, as it is likely biased low for diffuse gas associated with satellite subhaloes.




A more detailed assessment of this effect, including a reanalysis of the main gas-related tests in TNG300 and/or with updated subhalo-identification procedures, is left for future work. The comparison presented here is intended only to provide an approximate estimate of the magnitude of the bias and to guide the interpretation of the gas-related results reported in this paper.}
\end{appendix}
\end{document}